\documentclass[useAMS,usenatbib,usegraphicx]{mn2e}
\usepackage{amssymb,amsmath}
\newcommand{\msano}{{\rm M}_\odot ~{\rm yr}^{-1}}
\newcommand{\jdot}{\dot{J}}
\newcommand{\mdot}{\dot{M}}

\title[The stellar wind cycles and planetary radio emission of the $\tau$~Boo system]{The stellar wind cycles and planetary radio emission of the $\boldsymbol\tau$~Boo system}
\author[A.~A.~Vidotto et al.]{A.~A.~Vidotto$^{1}$\thanks{E-mail: Aline.Vidotto@st-andrews.ac.uk}, {R.~Fares}$^{1}$, {M.~Jardine}$^{1}$, {J.-F.~Donati}$^{2}$, {M.~Opher}$^{3}$, {C.~Moutou}$^{4}$, \newauthor  {C.~Catala}$^{5}$ and {T.~I.~Gombosi}$^{6}$ \\
$^{1}$SUPA, School of Physics and Astronomy, University of St Andrews, North Haugh, St Andrews, KY16 9SS, UK\\
$^{2}$LATT-UMR 5572, CNRS \& Univ. P. Sabatier, 14 Av.~E.~Belin, Toulouse, F-31400, France\\
$^{3}$Department of Astronomy, Boston University, 725 Commonwealth Avenue, Boston, MA, 02215, USA\\
$^{4}$LAM-UMR 6110, CNRS \& Univ. de Provence, 38 rue Fr\'ederic Juliot-Curie, Marseille, F-13013, France\\
$^{5}$LESIA-UMR 8109, CNRS \& Univ. Paris VII, 5 Place Janssen, Meudon-Cedex, F-92195, France\\
$^{6}$Dept.~of Atmospheric, Oceanic and Space Sciences, Univ.~Michigan, 1517 Space Research Building, Ann Arbor, MI, 48109, USA
}
\begin{document}

\date{Accepted . Received ; in original form }

\pagerange{\pageref{firstpage}--\pageref{lastpage}} \pubyear{2012}

\maketitle

\label{firstpage}

\begin{abstract}
$\tau$~Boo is an intriguing planet-host star that is believed to undergo magnetic cycles similar to the Sun, but with a duration that is about one order of magnitude smaller than that of the solar cycle. With the use of observationally derived surface magnetic field maps, we simulate the magnetic stellar wind of $\tau$~Boo by means of three-dimensional MHD numerical simulations. As the properties of the stellar wind depend on the particular characteristics of the stellar magnetic field, we show that the wind varies during the observed epochs of the cycle. Although the mass loss-rates we find ($\sim 2.7 \times 10^{-12}~\msano$) vary less than 3 per cent during the observed epochs of the cycle, our derived angular momentum loss-rates vary from $1.1$ to $2.2 \times 10^{32}$ erg. The spin-down times associated to magnetic braking range between $39$ and $78$~Gyr. We also compute the emission measure  from the (quiescent) closed corona and show that it remains approximately constant through these epochs at a value of $\sim 10^{50.6}~{\rm cm}^{-3}$. This suggests that a magnetic cycle of $\tau$~Boo may not be detected by X-ray observations. We further investigate the interaction between the stellar wind and the planet by estimating radio emission from the hot-Jupiter that orbits at $0.0462$~au from $\tau$~Boo. By adopting reasonable hypotheses, we show that, for a planet with a magnetic field similar to Jupiter ($\sim14$~G at the pole), the radio flux is estimated to be about $0.5-1$~mJy, occurring at a frequency  of $34$~MHz. If the planet is  less magnetised (field strengths roughly smaller than $4$~G), detection of radio emission from the ground is unfeasible due to the Earth's ionospheric cutoff. According to our estimates, if the planet is more magnetised than that and provided the emission beam crosses the observer line-of-sight, detection of radio emission from $\tau$~Boo b is only possible by ground-based instruments with a noise level of $\lesssim 1$~mJy, operating at low frequencies. 
\end{abstract}


\section{INTRODUCTION}
\subsection{Magnetic Cycles and Stellar Winds}
$\tau$~Boo (spectral type F7V)  is a remarkable object, not only because it hosts a giant planet orbiting very close to the star, but also because it is one of the few stars for which magnetic polarity reversals have been reported in the literature. So far, two polarity reversals have been detected \citep{2008MNRAS.385.1179D, 2009MNRAS.398.1383F}, suggesting that it undergoes magnetic cycles similar to the Sun, but with a period that is about one order of magnitude smaller than the solar one. The polarity reversals in $\tau$~Boo seem to occur at a period of roughly one year, switching from a negative poloidal field near the visible pole in June-2006 \citep{2007MNRAS.374L..42C} to a positive poloidal field in June-2007 \citep{2008MNRAS.385.1179D}, and then back again to a negative polarity in July-2008 \citep{2009MNRAS.398.1383F}. At these three observing epochs, $\tau$~Boo presented a dominant poloidal field, but in between the last observed reversal (more specifically, in January-2008), the magnetic field of $\tau$~Boo switched to a predominantly toroidal one. Subsequent observations confirm that $\tau$~Boo presents stable, periodic polarity reversals (Fares et al., in prep), confirming the presence of a magnetic cycle with a duration of roughly 2 years. 

It is interesting to note that polarity reversals have been observed in other objects, but the confirmation of the presence of a cycle requires a long-term monitoring. \citet{2009A&A...508L...9P} observed the first polarity switch, mostly visible in the azimuthal component of the magnetic field, in the solar-mass star HD~190771. Later on, \citet{2011AN....332..866M} observed another polarity switch of HD~190771. Contrary to $\tau$~Boo, the initial magnetic state of this object was not recovered, suggesting that the polarity reversals in HD~190771 does not take the form of a solar-type cycle. Polarity reversals have also been reported in the other solar-type stars HD 78366 and $\xi$~Boo~A  \citep{2011AN....332..866M}, while the young star  HR~1817 showed an ``attempted'' reversal in the azimuthal component of the magnetic field \citep{2010IAUS..264..130M}. In this case, rather than undergoing a reversal, the magnetic field strength decreased, but then strengthened with the same polarity \citep{2010IAUS..264..130M}.

The nature of such a short magnetic cycle in $\tau$~Boo remains an open question. Differential rotation is thought to play an important role in the solar cycle. The fact that $\tau$~Boo presents a much higher level of surface differential rotation than that of the Sun may be responsible for its short observed cycle\footnote{Note however that, although also known to present high levels of differential rotation, HD~171488 seems not to have a fast magnetic cycle as the one reported for $\tau$~Boo \citep{2006MNRAS.370..468M, 2011MNRAS.411.1301J}.}. 
 In addition, $\tau$~Boo also hosts a close-in planet that, due to its close proximity to the star, may have been able to synchronise, through tidal interactions, the rotation of the shallow convective envelope of the host star with the planetary orbital motion. This presumed synchronisation may enhance the shear at the tachocline, which may influence the magnetic cycle of the star \citep{2009MNRAS.398.1383F}. 

As the stellar winds of cool stars are magnetic by nature, variations of the stellar magnetic field during the cycle directly influences the outflowing wind. The solar wind, for instance, is dominated by high-speed flows outflowing from the coronal holes during and near the minimum phases of the solar cycle. As the solar cycle approaches its maximum, the coronal holes become smaller and the high-speed flows narrow and weaken \citep{2007bsw..book.....M}. In analogy to the Sun, we expect that the stellar wind from $\tau$~Boo will respond to variations in the magnetic properties of the star during its cycle. One of the goals of this present study is to quantify these variations.

\subsection{Radio Emission from Wind-Planet Interaction}\label{intro.radio}
The solar magnetic cycle has a direct impact on the planets of the Solar System. In particular at the Earth, during periods of intense solar activity, geomagnetic storms can, e.g., produce auroras, disrupt radio transmissions, affect power grids, and damage satellites orbiting the Earth. Likewise, the magnetic cycle of $\tau$~Boo should affect any orbiting planet, especially if located at such a close distance as that of $\tau$~Boo b \citep[$0.0462~{\rm au} \simeq 6.8~R_\star$ from its host-star,][]{butler}.

In particular, the planet's interaction with the host star wind may lead to planetary radio emission. Radio emission has been detected in the giant planets of the solar system, in the Earth and in a few satellites orbiting these planets \citep{1998JGR...10320159Z}. The auroral radio emission from the Earth, for instance, is pumped (primarily) by reconnection events between the interplanetary magnetic field (embedded in the solar wind) and the planet's own magnetic field at the dayside magnetopause. On the other hand, the interaction of Jupiter with its moon Io, which also generates radio emission, is thought to be caused by the satellite's motion inside the planet's magnetosphere \citep{1980JGR....85.1171N}. 

Radio emission from exoplanets has been investigated by several authors \citep[e.g., ][]{2000ApJ...545.1058B,2001Ap&SS.277..293Z,2004ApJ...612..511L,2005MNRAS.356.1053S,2005A&A...437..717G,2007P&SS...55..618G,2007A&A...475..359G,2007P&SS...55..598Z,2007ApJ...668.1182L,2008A&A...490..843J,2010AJ....139...96L,2010AJ....140.1929L,2010ApJ...720.1262V,2011MNRAS.414.1573V,2011MNRAS.414.2125N}. Detection of the auroral radio signatures from exoplanets would consist of a direct planet-detection method, as opposed to the widely used indirect methods of radial velocity measurements or transit events. Moreover, the detection of exoplanetary radio emission would comprise a way to assess the magnitude of planetary magnetic fields.   

However, despite many attempts, radio emission from exoplanets has not been detected so far. One of the reasons for the unsuccessful detection is attributed to the beamed nature of the electron-cyclotron maser instability, believed to be the process operating in the generation of radio emission. Poor instrumental sensitivity is also pointed to as an explanation for the lack of detection of radio emission from exoplanets. Another reason for the failure is often attributed to frequency mismatch: the emission process is thought to occur at cyclotron frequencies, which depend on the intensity of the planetary magnetic field. Therefore, planets with magnetic field strengths of, e.g., a few G would emit at a frequency that could be either unobserved from the ground due to the Earth's ionospheric cut-off or that does not correspond to the operating frequencies of available instruments. In that regard, the low-operating frequency of LOFAR (current under commissioning), jointly with its high sensitivity at this low-frequency range, makes it an instrument that has the potential to detect radio emission from exoplanets. A more thorough discussion about the non-detection of radio emission from exoplanets can be found in, e.g., \citet{2000ApJ...545.1058B, 2010AJ....140.1929L}.

\subsection{This Work}
To quantify the effect the stellar cycle has on the orbiting planet, one has to understand the properties of the stellar wind, which depends on the particular geometry of the coronal magnetic field at each epoch during the stellar cycle. Several works have studied the influence of the geometry of the coronal magnetic field on the stellar wind properties  by means of numerical simulations \citep[][among others]{1999A&A...343..251K, 2000ApJ...530.1036K, 2008ApJ...678.1109M, 2009ApJ...699..441V, 2009ApJ...703.1734V, 2010ApJ...720.1262V, 2011MNRAS.412..351V, 2010ApJ...721...80C, 2011ApJ...737...72P}. Here, we implement the observationally-derived surface magnetic field of $\tau$~Boo in our numerical model. In total, we use in the present study four surface magnetic maps derived at four different epochs: June-2006, June-2007, January-2008, and July-2008. These maps  have been presented elsewhere \citep{2007MNRAS.374L..42C, 2008MNRAS.385.1179D, 2009MNRAS.398.1383F} and encompass (at least) one full cycle, with two polarity reversals in the poloidal field.

Using the results of our stellar wind models, we then evaluate the radio flux emitted by $\tau$~Boo~b, analogously to what is observed for the giant planets in the solar system. The radio emission is calculated at each observed epoch of the stellar magnetic cycle. 
We note that the $\tau$~Boo system has been classified among the prime targets for radio emission detection based on models that consider more simplistic descriptions for the stellar wind \citep[e.g.,][]{2004ApJ...612..511L, 2005MNRAS.356.1053S, 2008A&A...490..843J, 2010A&A...522A..13R, 2011RaSc...46.0F09G}. A more sophisticated model for the stellar wind, such as the one presented in this paper, will be useful to provide some interpretation of the radio emission when it is discovered.

This paper is organised as follows: Section~\ref{sec.wind} presents the three-dimensional numerical model used in our simulations and describes the observed surface magnetic field distributions that are implemented in our model. Section~\ref{sec.results_wind} presents the results of our stellar wind modelling, and in Section~\ref{sec.radio}, we investigate planetary radio emission arising from the interaction between the stellar wind and the planet. Section~\ref{sec.conc} presents further discussion and the conclusions of our work.

\section{Stellar Wind Model}\label{sec.wind}
\subsection{Numerical Model}
The lack of symmetry in the magnetic field distribution at the surface of $\tau$~Boo requires the stellar wind equations to be solved in a fully three-dimensional geometry. Our simulations make use of the three-dimensional MHD numerical code BATS-R-US developed at University of Michigan \citep{1999JCoPh.154..284P}. BATS-R-US has been widely used to simulate, e.g., the Earth's magnetosphere \citep{2006AdSpR..38..263R}, the heliosphere \citep{2003ApJ...595L..57R}, the outer-heliosphere \citep{1998JGR...103.1889L, 2003ApJ...591L..61O, 2004ApJ...611..575O}, coronal mass ejections \citep{2004JGRA..10901102M,2005ApJ...627.1019L}, and the magnetosphere of planets \citep{2004JGRA..10911210T,2005GeoRL..3220S06H}, among others. It solves the ideal MHD equations, that in the conservative form are given by
\begin{equation}
\label{eq:continuity_conserve}
\frac{\partial \rho}{\partial t} + \boldsymbol\nabla\cdot \left(\rho {\bf u}\right) = 0,
\end{equation}
\begin{equation}
\label{eq:momentum_conserve}
\frac{\partial \left(\rho {\bf u}\right)}{\partial t} + \boldsymbol\nabla\cdot\left[ \rho{\bf u\,u}+ \left(p + \frac{B^2}{8\pi}\right)I - \frac{{\bf B\,B}}{4\pi}\right] = \rho {\bf g},
\end{equation}
\begin{equation}
\label{eq:bfield_conserve}
\frac{\partial {\bf B}}{\partial t} + \boldsymbol\nabla\cdot\left({\bf u\,B} - {\bf B\,u}\right) = 0,
\end{equation}
\begin{equation}
\label{eq:energy_conserve}
\frac{\partial\varepsilon}{\partial t} +  \boldsymbol\nabla \cdot \left[ {\bf u} \left( \varepsilon + p + \frac{B^2}{8\pi} \right) - \frac{\left({\bf u}\cdot{\bf B}\right) {\bf B}}{4\pi}\right] = \rho {\bf g}\cdot {\bf u} ,
\end{equation}
where the eight primary variables are the mass density $\rho$, the plasma velocity ${\bf u}=\{ u_r, u_\theta, u_\varphi\}$, the magnetic field ${\bf B}=\{ B_r, B_\theta, B_\varphi\}$, and the gas pressure $p$. The gravitational acceleration due to the star with mass $M_\star$ and radius $R_\star$ is given by ${\bf g}$, and $\varepsilon$ is the total energy density given by 
\begin{equation}\label{eq:energy_density}
\varepsilon=\frac{\rho u^2}{2}+\frac{p}{\gamma-1}+\frac{B^2}{8\pi} .
\end{equation}
We consider an ideal gas, so $p=n k_B T$, where  $k_B$ is the Boltzmann constant, $T$ is the temperature, $n=\rho/(\mu m_p)$ is the particle number density of the stellar wind, $\mu m_p$ is the mean mass of the particle, and $\gamma$ is the polytropic index (such that $p \propto \rho^\gamma$).

\begin{figure*}
\includegraphics[width=84mm]{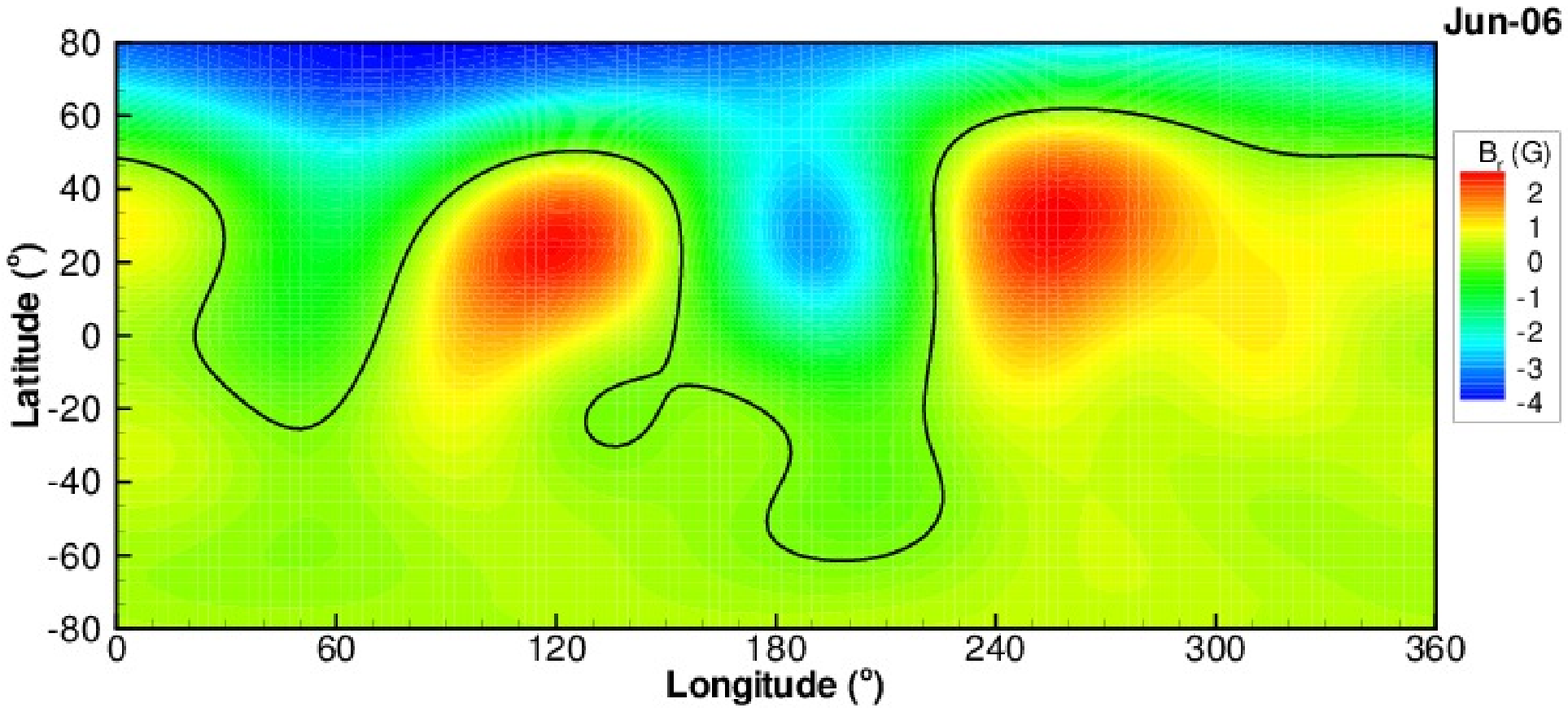}
\includegraphics[width=84mm]{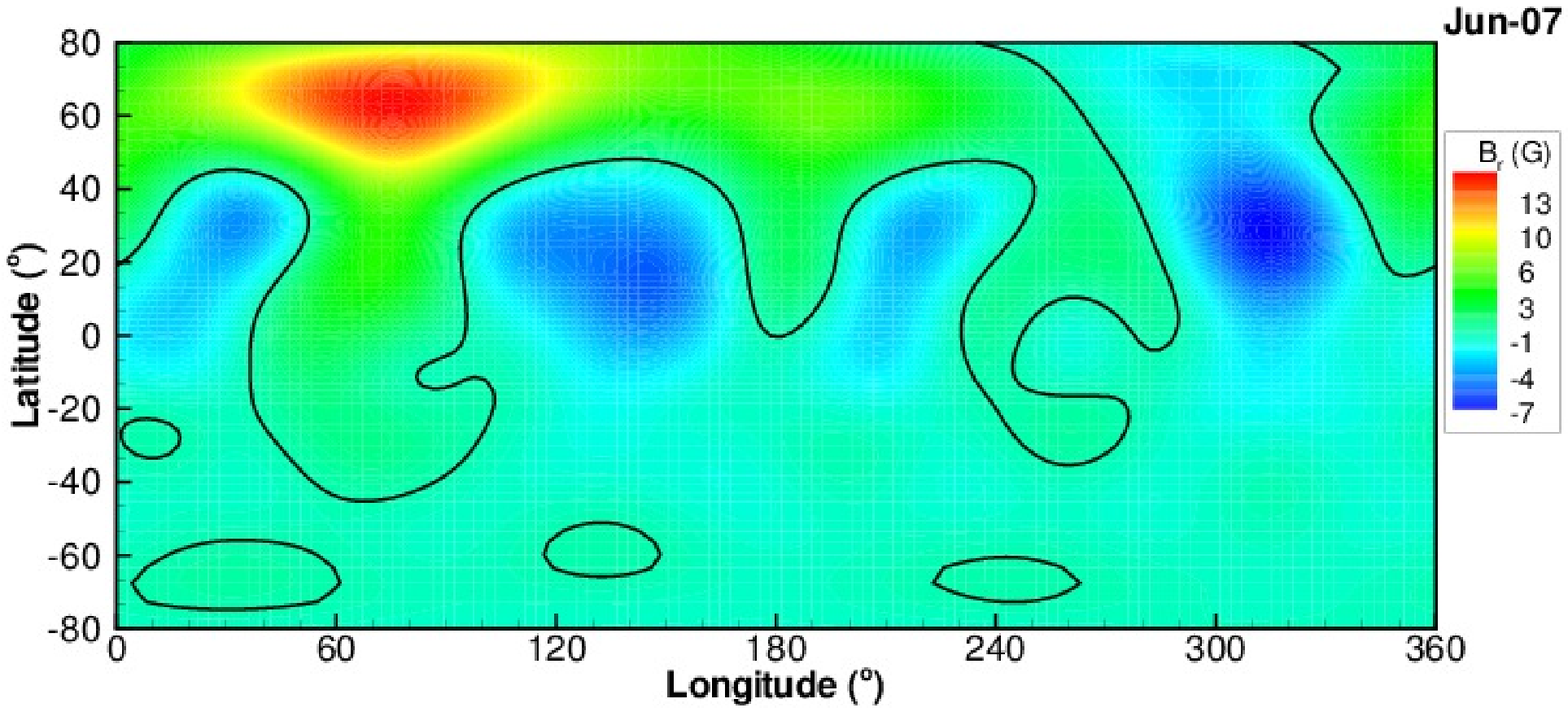}\\
\includegraphics[width=84mm]{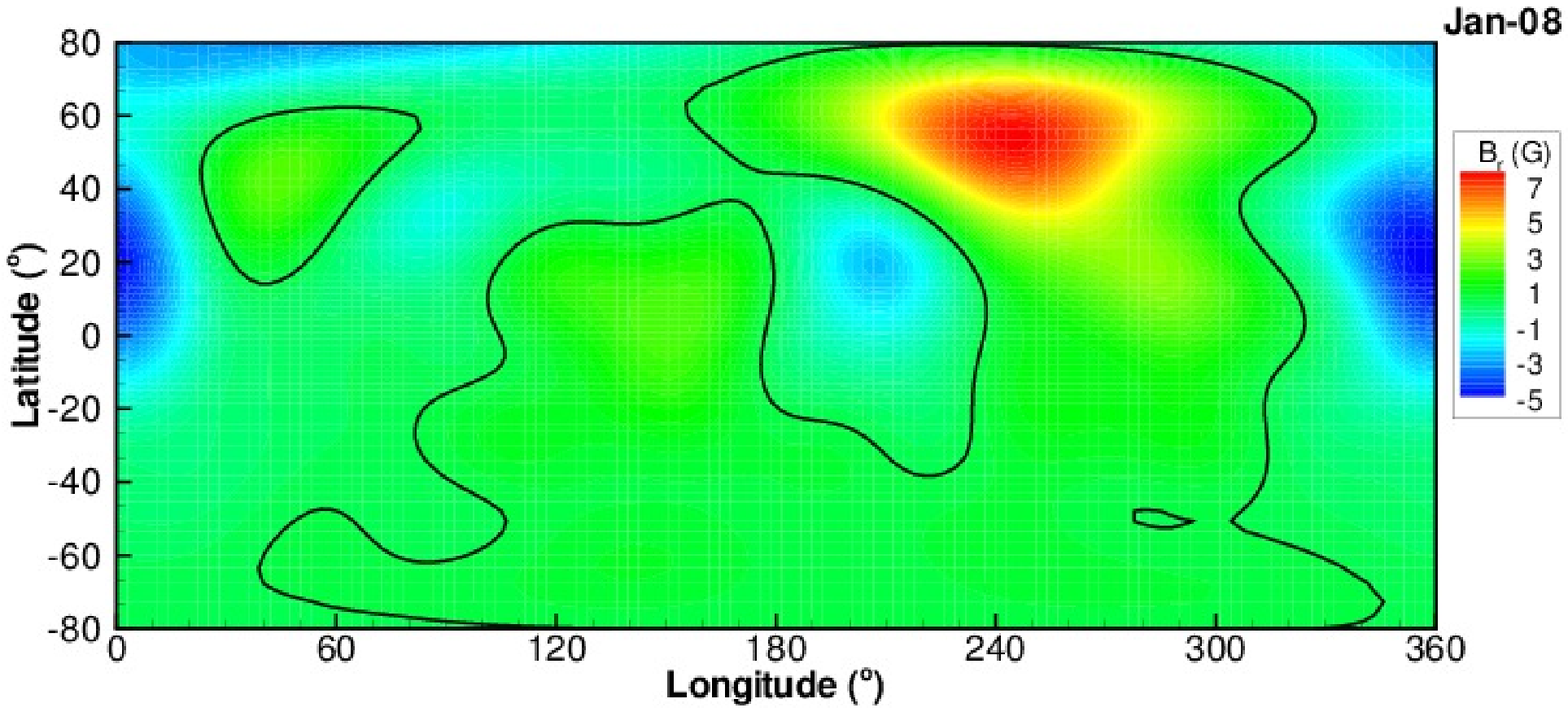}
\includegraphics[width=84mm]{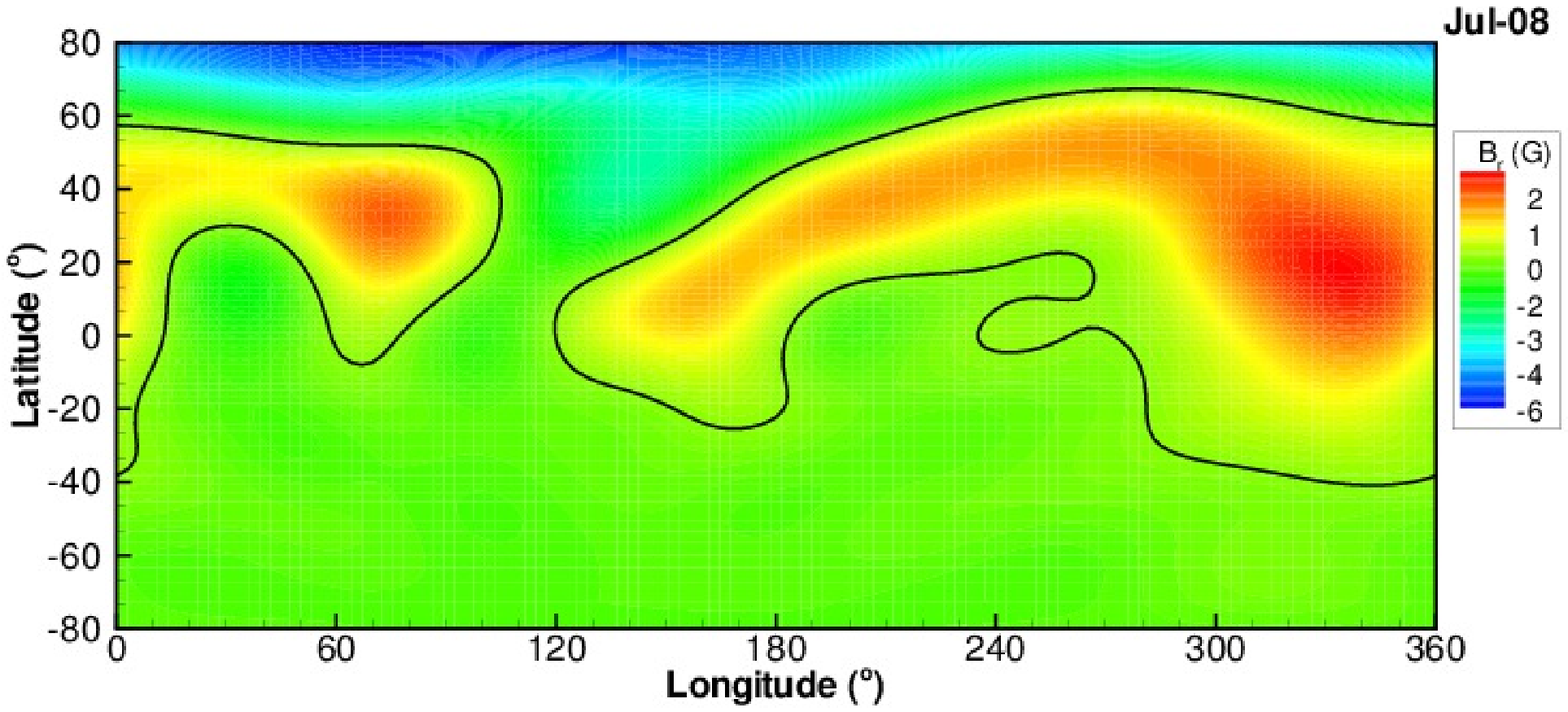}
\caption{Surface distribution of the radial component of the magnetic field of $\tau$~Boo reconstructed from observations using ZDI \citep{2007MNRAS.374L..42C,2008MNRAS.385.1179D, 2009MNRAS.398.1383F}. The black solid lines represent $B_r=0$~G. \label{fig.magnetograms}}
\end{figure*}

At the initial state of the simulations, we assume that the wind is thermally driven \citep{1958ApJ...128..664P}. At the base of the corona ($r=R_\star$), we adopt a wind coronal temperature $T_0$, wind number density $n_0$, and stellar rotation period $P_{\rm rot}$. The values adopted in our simulations are shown in Table~\ref{table_par}. The values of $M_\star$ and $R_\star$ are from \citet{2007ApJS..168..297T} and $P_{\rm rot}$ from \citet{2009MNRAS.398.1383F}. The value of $T_0$ we adopted represents a typical temperature of a stellar corona and $n_0$ is selected in such a way as to recover observed emission measure values (see Section~\ref{sec.results_wind}). With this numerical setting, the initial solution for the density, pressure (or temperature) and wind velocity profiles are fully specified. 

\begin{table} 
\centering
\caption{Adopted parameters for the simulations. \label{table_par}}    
\begin{tabular}{lll}  
\hline
&Parameter	&	Value	 \\ \hline
Stellar mass &$M_\star ~(M_\odot)$  & $   1.341   $ \\
Stellar radius&$R_\star ~(R_\odot)$  & $   1.46   $ \\
Stellar rotation period&$P_{\rm rot}$ (d)  & $   3.0   $ \\
Coronal base temperature&$T_0$ (MK)  & $   2   $ \\
Coronal base density &$n_0 ~({\rm cm}^{-3})$  & $   10^9   $ \\
Polytropic index&$\gamma$  & $   1.1   $ \\
Particle mean mass&$\mu~(m_p)$  & $  0.5  $ \\
 \hline
\end{tabular}
\end{table}

To complete our initial numerical set up, we assume that the magnetic field is potential everywhere (i.e., $\boldsymbol\nabla \times {\bf B}=0$). To provide an initial solution for ${\bf B}$, we use the potential field source surface method \citep[PFSSM,][]{1969SoPh....9..131A, 2002MNRAS.333..339J}, which assumes that beyond a given radius (which defines a spherical {\it source surface}), the magnetic field lines are purely radial. The initial solution for ${\bf B}$ is found once the radial component of the magnetic field $B_r$ at the surface of the star is specified and a distance to the source surface is assumed (set at $4~R_\star$ in the initial state of our runs).  In our simulations, we incorporate $B_r$ derived from the observations, similarly to the method presented in \citet{2011MNRAS.412..351V}. Section~\ref{sec.maps} presents the surface magnetic field maps used in this study.

Once set at the initial state of the simulation, the distribution of $B_r$ is held fixed at the surface of the star throughout the simulation run, as are the coronal base density and thermal pressure. A zero radial gradient is set to the remaining components of ${\bf B}$ and ${\bf u}=0$ in the frame corotating with the star. The outer boundaries at the edges of the grid have outflow conditions, i.e., a zero gradient is set to all the primary variables. The rotation axis of the star is aligned with the $z$-axis, and the star is assumed to rotate as a solid body.

Our grid is Cartesian and extends in $x$, $y$, and $z$ from $-20$ to $20~R_\star$, with the star placed at the origin of the grid. BATS-R-US uses block adaptive mesh refinement (AMR), which allows for variation in numerical resolution within the computational domain. The finest resolved cells are located close to the star (for $r \lesssim 4~R_\star$), where the linear size of the cubic cell is $0.039~R_\star$. The coarsest cell is about one order of magnitude larger (linear size of $0.31~R_\star$) and is located at the outer edges of the grid. The total number of cells in our simulations is about $6.5$ million. 

As the simulations evolve in time, both the wind and magnetic field lines are allowed to interact with each other. The resultant solution, obtained self-consistently, is found when the system reaches steady state (in the reference frame corotating with the star).  Our simulations run for two to three stellar rotations periods ($6$ to $9$ days of physical time). Despite the initial assumption of a potential field, we remind the reader that the steady-state solution, shown in the remainder of this paper, deviates from a potential solution and currents are created in the system (see Appendix~\ref{appendix} for a comparison between both solutions). Likewise, the initially spherically symmetric hydrodynamical quantities ($\rho$, $p$, ${\bf u}$) evolve to asymmetric distributions.

\subsection{Adopted Surface Magnetic Field Distributions}\label{sec.maps}
The surface magnetic maps used in this study were reconstructed using Zeeman-Doppler Imaging (ZDI), a tomographic imaging technique \citep[e.g.,][]{1997A&A...326.1135D}. Using ZDI, one can reconstruct the large-scale magnetic field (intensity and orientation) at the surface of the star from a series of circular polarisation spectra. The radial component of the reconstructed surface magnetic maps are shown in Figure~\ref{fig.magnetograms} for the four different epochs considered here: June-2006 \citep{2007MNRAS.374L..42C}, June-2007 \citep{2008MNRAS.385.1179D}, January-2008 and July-2008 \citep{2009MNRAS.398.1383F}. 

\begin{figure}
\includegraphics[width=84mm]{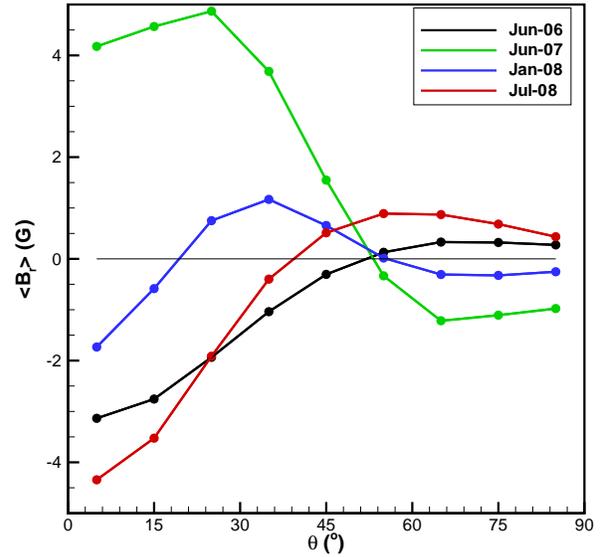}
\caption{Average intensity of $B_r$ calculated at bins of colatitude $\theta$. Along with Table~\ref{table_B} and Figure~\ref{fig.magnetograms}, it shows the similarities between the maps obtained in June-2006 (black solid line) and July-2008 (red solid line). \label{fig.average_colatitude}}
\end{figure}

Table~\ref{table_B} presents a summary of the main properties of the observed large-scale magnetic field distributions. The unsigned surface magnetic flux is calculated over the surface of the star ($S_\star$) as $\Pi_0 = \oint |B_r (R_\star)| {\rm d}S_\star$. 
Figure~\ref{fig.average_colatitude} shows the average value of the radial component calculated over colatitude bins of $10^{\rm o}$.
From Table~\ref{table_B} and Figures~\ref{fig.magnetograms} and \ref{fig.average_colatitude}, we note that the magnetic field distribution in July-2008 shows similarities to the one in June-2006, suggesting that the magnetic state of June-2006 seems to be recovered in July-2008. \citet{2009MNRAS.398.1383F} found a preferred cycle period of about $P_{\rm cyc}=800$~d, although a much shorter period of $250$~d was not excluded. For $P_{\rm cyc}=800$~d, the maps derived in June-2006 and July-2008 should describe similar phases at two {consecutive} cycles. 

\begin{table} 
\centering
\caption{Main properties of the observed surface magnetic field. The polarity of the poloidal component is described around the visible rotation pole and the phase of the cycle is calculated assuming a cycle period of $P_{\rm cyc} = 800$~days. The unsigned surface magnetic flux in the radial component is $\Pi_0$.  \label{table_B}}    
\begin{tabular}{ccccc}  
\hline
Date & Cycle & Dominant & Polarity & $\Pi_0$\\
& phase & component &  & ($10^{22}$~Mx) \\
\hline
Jun-06 & 0.12 & poloidal & negative & $11.4 $ \\
Jun-07 & 0.58 & poloidal & positive & $23.5 $ \\
Jan-08 & 0.86 & toroidal & mixed & $14.0 $ \\
Jul-08 & 0.07 & poloidal & negative & $12.1 $ \\
 \hline 
\end{tabular}
\end{table}

We note that, due to the lack of information in the unseen hemisphere of the star (latitudes $\lesssim-40^{\rm o}$), the reconstructed magnetic field there has essentially no energy (unconstrained field reconstruction). However, constraints on the property of the magnetic field in the unseen hemisphere could have been imposed \citep[e.g., see ][]{lang2012}. These constraints are adopted based on physical properties of the star. For example, for the classical T Tauri star BP Tau, anti-symmetric field configurations (with respect to the centre of the star) are preferred as they are the only ones capable of yielding the high-latitude anchoring of accretion funnels \citep{2008MNRAS.386.1234D}. Without a physical reason to justify the choice of a symmetric/anti-symmetric topology, we adopt in this work the unconstrained maps. 
In Appendix~\ref{sec.assumptions}, we show that different assumptions adopted during the field reconstruction in the unseen hemisphere of the star (latitudes $\lesssim-40^{\rm o}$) do not have an appreciable effect on the results presented in this paper.

\section{Results: Stellar Wind Properties}\label{sec.results_wind}
The properties of the stellar wind depend on the particular geometry of the coronal magnetic field, which varies through the stellar cycle. Ideally, if one wishes to investigate the {smooth} evolution of the stellar wind through the cycle, the evolution of the large-scale magnetic field at the base of the corona should be incorporated in the simulations by means of a time-dependent boundary condition for ${\bf B}$. To do that, one needs surface magnetic distributions that are reasonably well time-sampled throughout the cycle. 

In the present study, we make use of four available surface magnetic maps that were obtained at intervals of six months to one year. Therefore, the wind solutions found for each of these observing epochs represent a {\it snapshot} of the stellar wind at each given epoch. Implicitly, we are assuming that the time for the wind to adjust to the evolution of the surface magnetic field occurs in a faster timescale than the dynamical timescale of the magnetic evolution. This sounds a reasonable hypothesis as, in our simulations, for a typical wind velocity of about $300~{\rm km~s}^{-1}$, a spatial scale of $20~R_\star$ (half-size of our grid) will be covered in less than a day, while, due to high surface differential rotation, it is likely that significant magnetic field evolution should occur on a longer time-scale of a few weeks\footnote{For a latitude rotational shear of $d\Omega = 0.46~{\rm rad~d}^{-1}$ \citep{2009MNRAS.398.1383F}, the time for the equator to lap the pole by one complete rotation cycle is about 2 weeks. }

\begin{figure*}
\includegraphics[width=84mm]{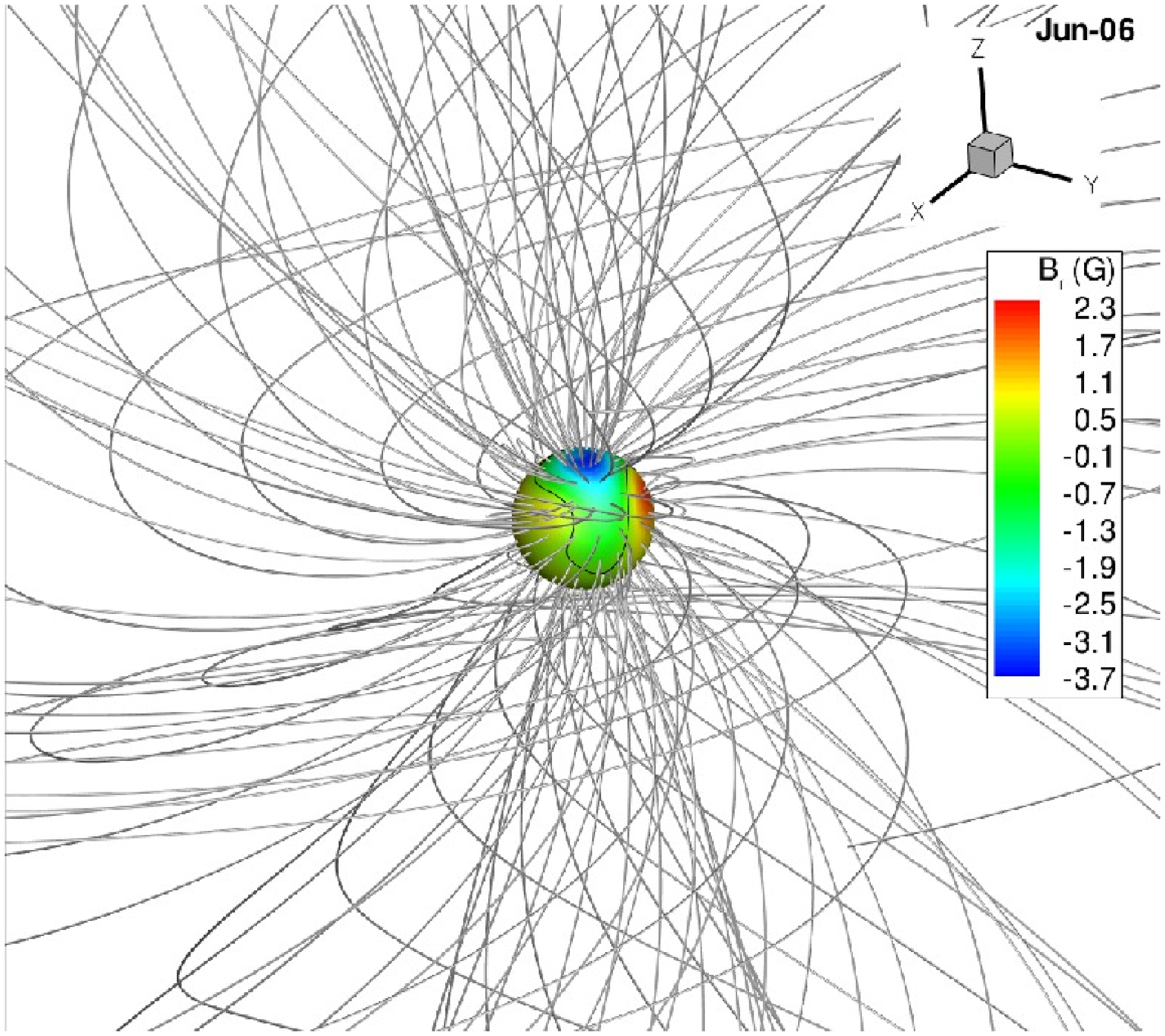}
\includegraphics[width=84mm]{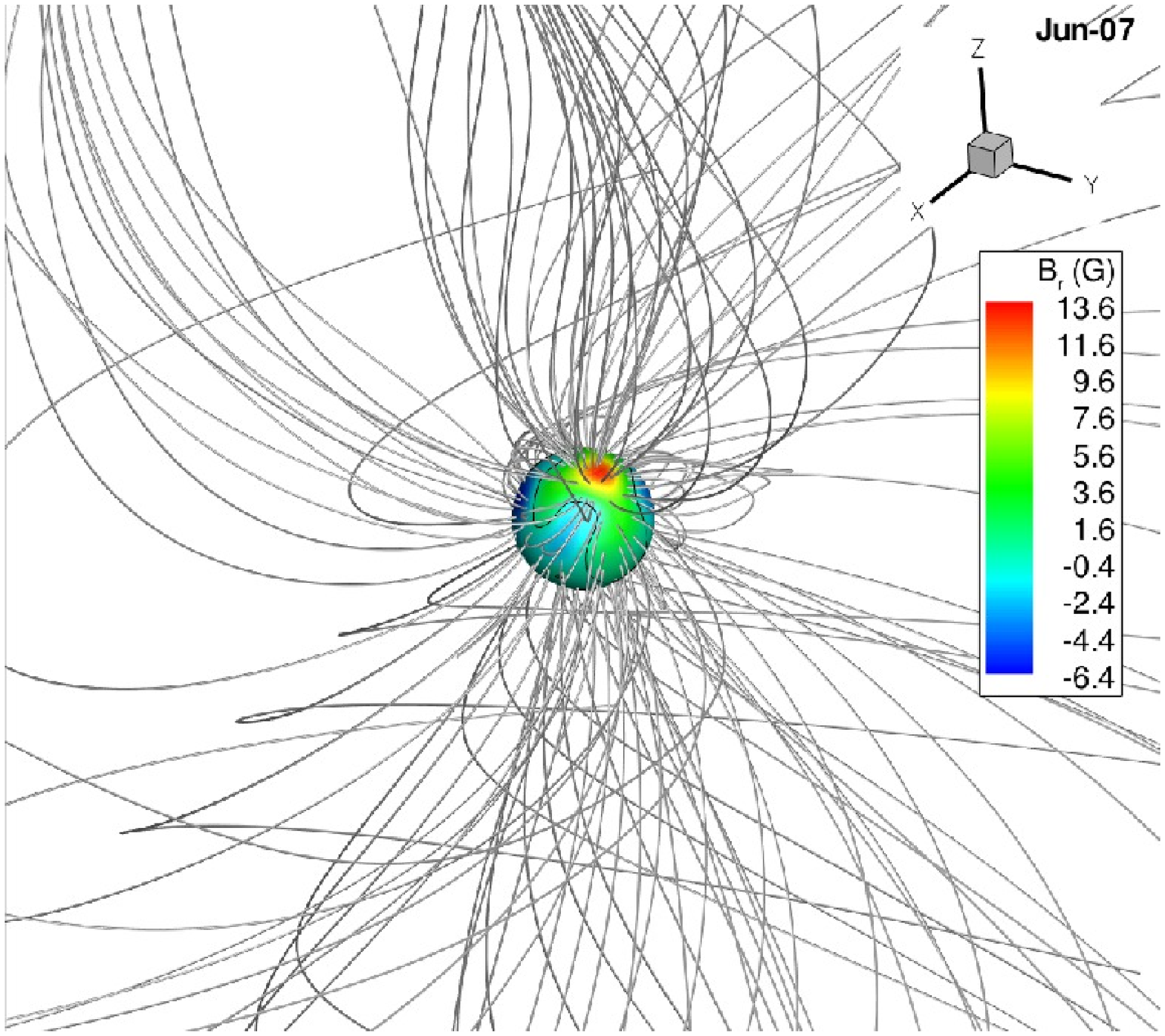}\\
\includegraphics[width=84mm]{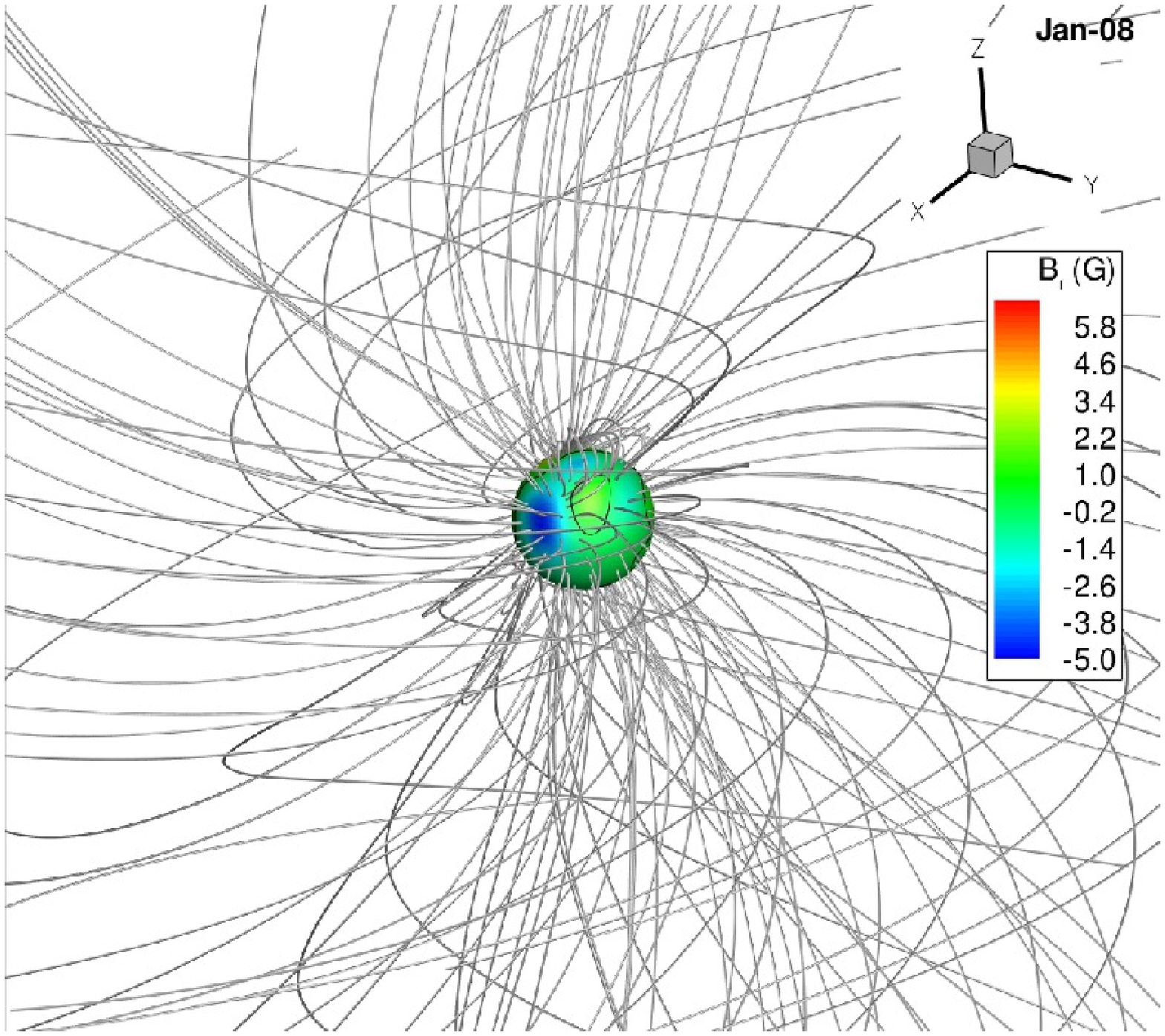}
\includegraphics[width=84mm]{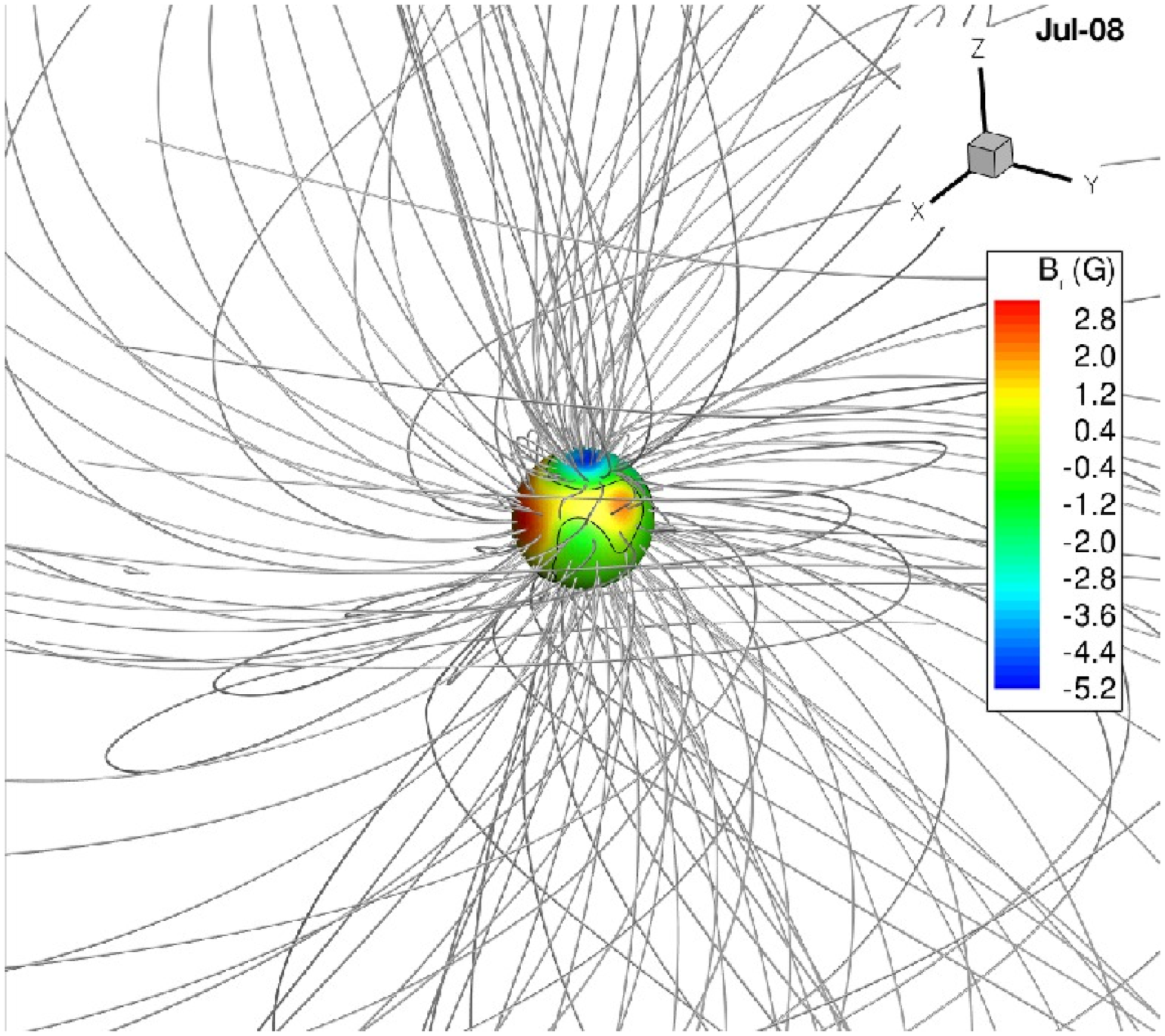}
\caption{Three-dimensional view of the final configuration of the magnetic field lines (grey lines) of the corona of $\tau$~Boo for the four cases analysed in this paper. The radial magnetic field are shown at the surface of the star in colour-scale. The rotational axis of the star is along the $z$-axis, the equator is in the $xy$-plane. \label{fig.Blines}}
\end{figure*}

\begin{figure*}
\includegraphics[width=84mm]{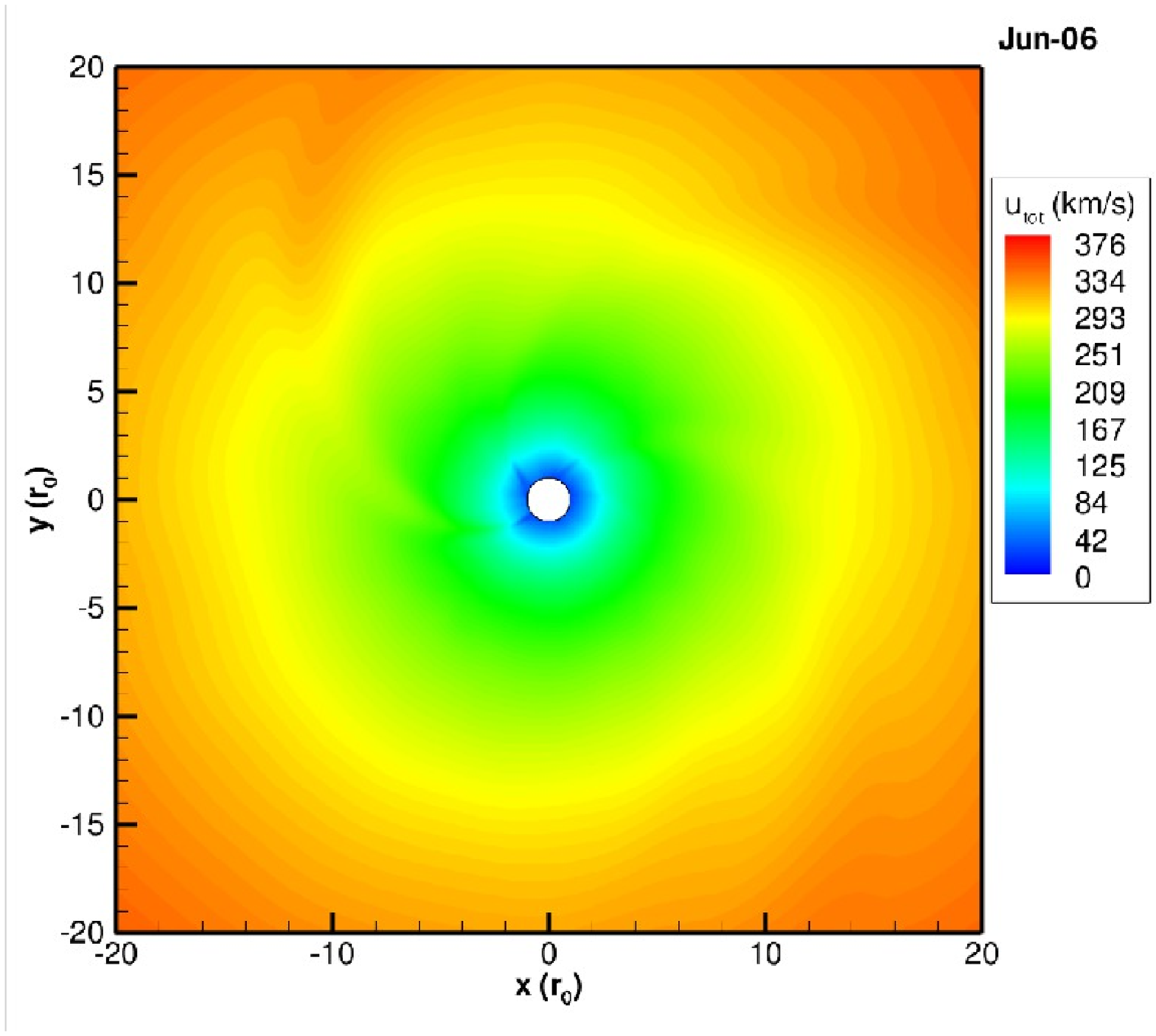}
\includegraphics[width=84mm]{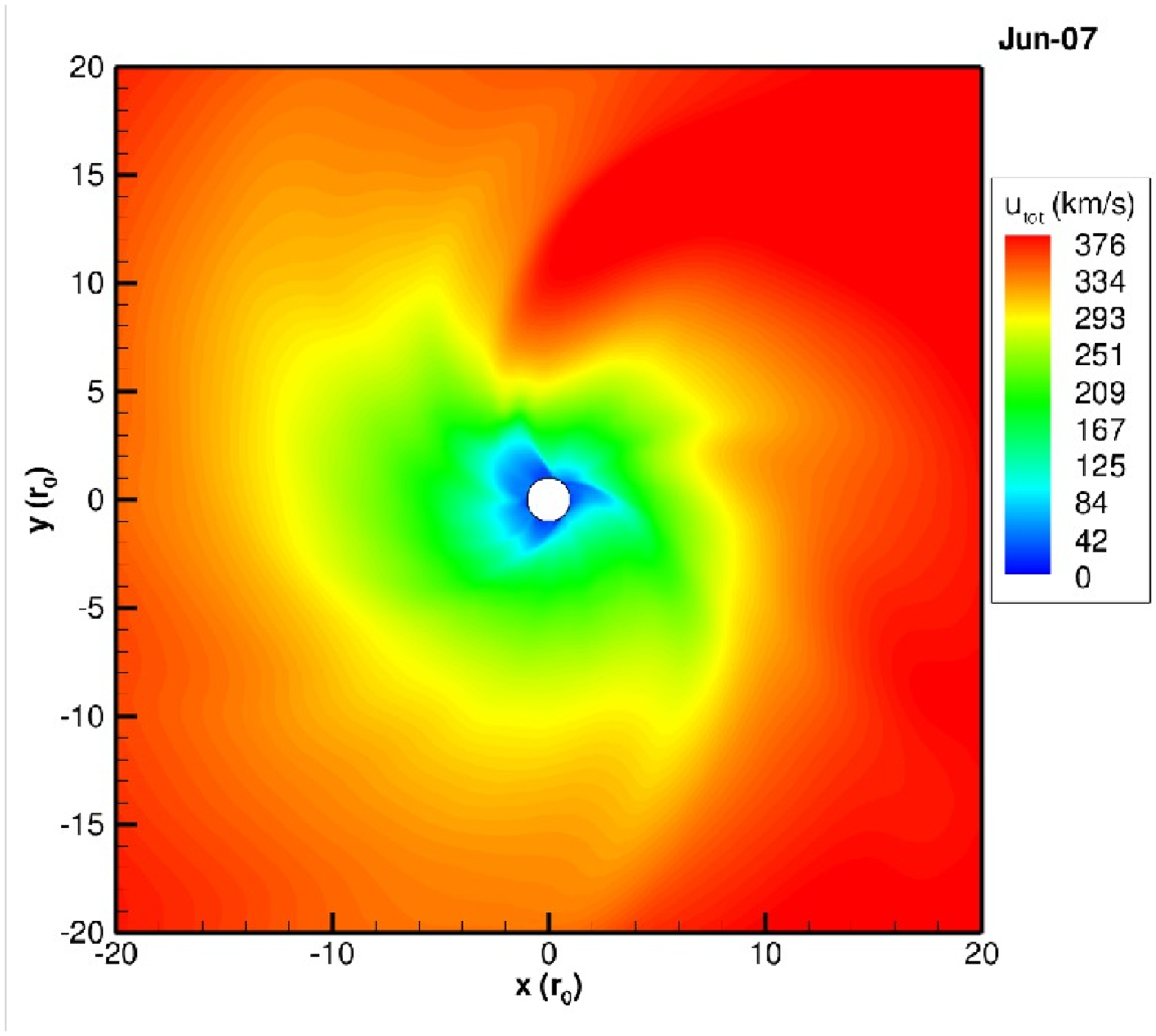}\\
\includegraphics[width=84mm]{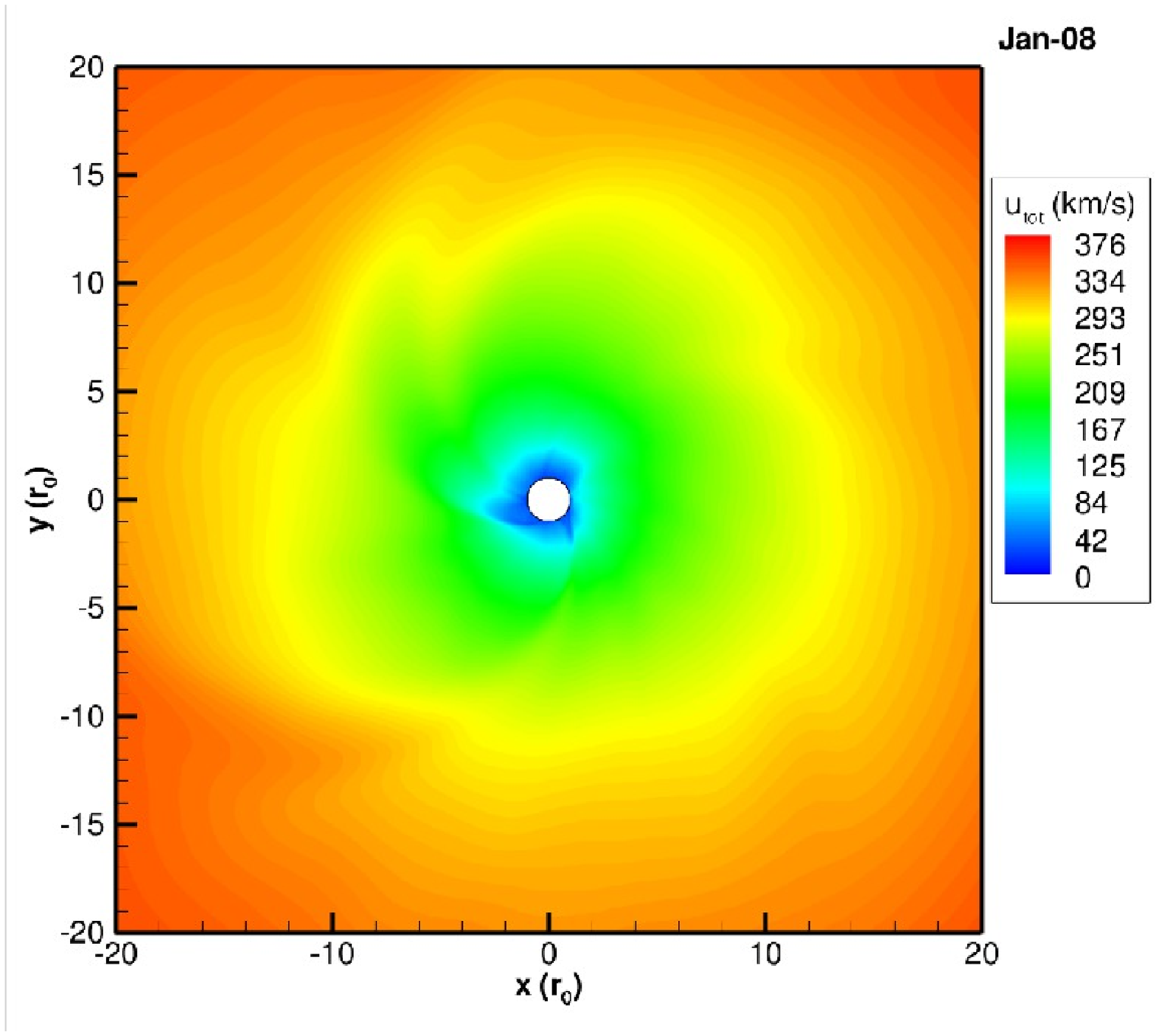}
\includegraphics[width=84mm]{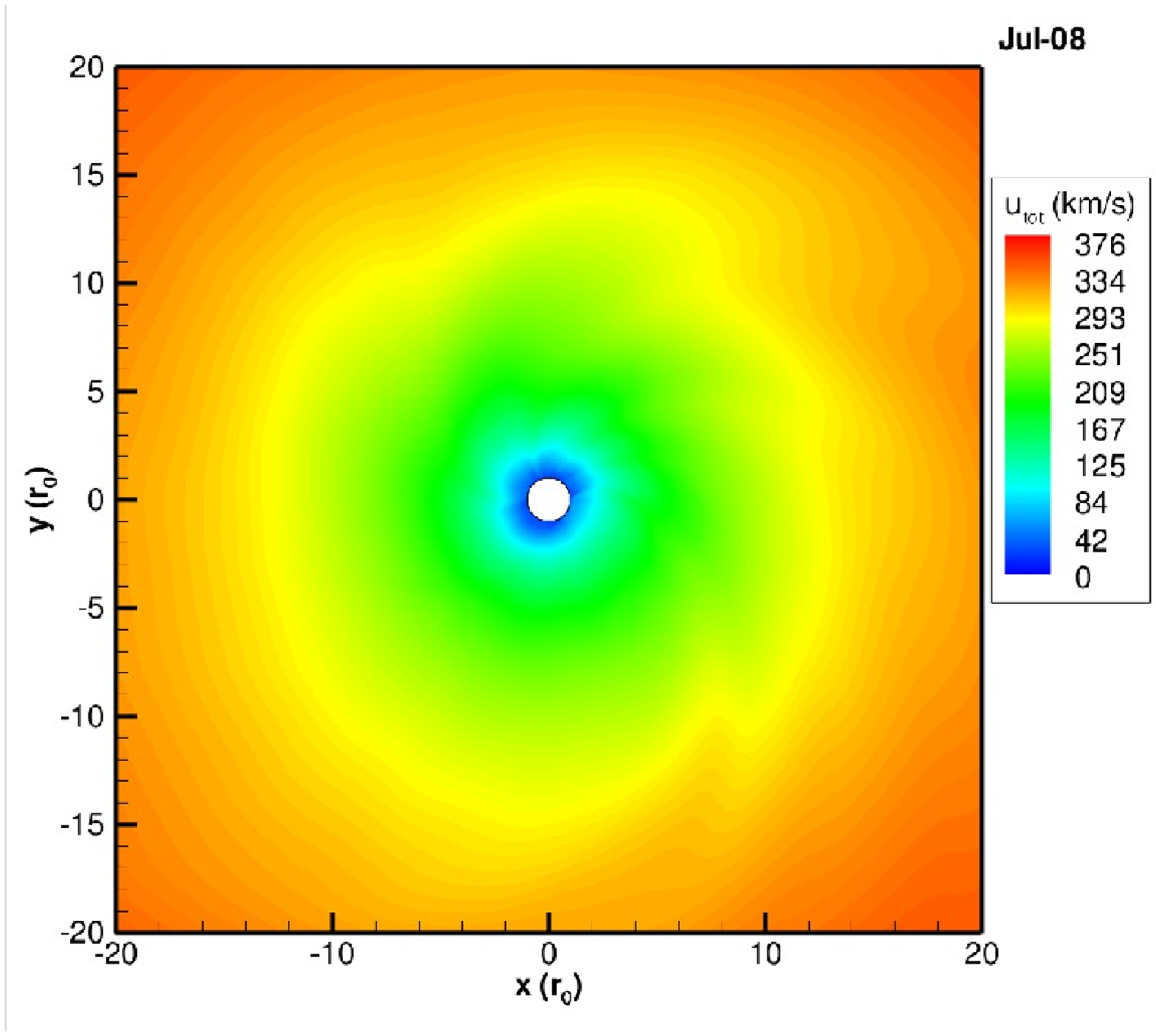}
\caption{The wind velocity distributions for each epoch are shown in the equatorial plane of the star. \label{fig.V}}
\end{figure*}

Figure~\ref{fig.Blines} shows the final configuration of the magnetic field lines in the corona of $\tau$~Boo for different epochs. We note that, due to the presence of the wind, the magnetic field lines become twisted around the rotation axis ($z$-axis). The wind velocity in the equatorial plane of the star ($xy$-plane) is shown in Figure~\ref{fig.V}. We note that both the coronal magnetic field lines and the wind velocity profile vary through the cycle. 

For each epoch, we compute the mass-loss rate, defined as the flux of particles flowing across a closed surface $S$ 
\begin{equation}
\dot{M} = \oint_S \rho {\bf u} \cdot {\rm d} {\bf S}.
\end{equation}
The calculated $\dot{M}$ for each cycle phase is presented in Table~\ref{table}, where we note that the derived mass-loss rates are about two orders of magnitude larger than the solar wind value ($\dot{M}_\odot \sim 2 \times 10^{-14}~\msano$). We warn the reader that the values of $\mdot$ obtained here (and, in general, by any stellar wind models) strongly depend on the choice of the base density $n_0$. One way to constrain coronal base densities is to perform a direct comparison between our derived mass-loss rates and the observationally determined ones. However, to the best of our knowledge, mass-loss rates determination for $\tau$~Boo are not available in the literature. 
\begin{table*} 
\centering
\caption{Summary of the results of the simulations. The columns are, respectively: the date of the observations from wihch the surface magnetic maps were derived (Figure~\ref{fig.magnetograms}), the mass-loss rate ($\mdot$), emission measure (EM) computed in the closed field-line region, angular momentum-loss rate ($\jdot$), spin-down time ($\tau$), unsigned open magnetic flux ($\Pi_{\rm open}$), and the ratio of the unsigned open magnetic flux to unsigned surface flux ($f_{\rm open}= \Pi_{\rm open}/\Pi_0$).
\label{table}}    
\begin{tabular}{ccccccc}  
\hline
Date&	$\dot{M} $	& EM &	$	\dot{J}	$	& $\tau$ &		$\Pi_{\rm open}$ & $f_{\rm open}$ \\
&	$	 (10^{-12} \msano)$ & $(10^{50} ~{\rm cm^{-3}})$ &	$ (10^{32} ~\rm{erg})$ & (Gyr)	&	$(10^{22} ~{\rm Mx})$ &   \\ \hline
Jun-06  &$    2.67$ &$4.4$& $1.2$ & $71$&$    9.5$ &  $0.84$ \\
Jun-07  &$    2.75$ &$4.3$& $2.2$ & $39$&$   20.5$ & $0.87$  \\
Jan-08  &$    2.69$ &$4.3$& $1.4$ & $61$&$   11.2$ & $ 0.80$ \\
Jul-08  &$    2.68$ &$4.4$& $1.1$ & $78$&$    9.0$ & $0.74$  \\
 \hline
\end{tabular}
\end{table*}
A less-direct way to constrain coronal base densities is through the comparison of emission measure (EM) values derived from X-ray spectra. Coronal X-ray emission comes from flaring loops with different sizes. The net effect of the superposition of the small-scale loops should be to form the observed regions of closed magnetic field lines (large-scale structure). Therefore, to compute the EM, we concentrate only on the closed-field line regions. The EM is defined as
\begin{equation}
{\rm EM} = \int n_e n_i {\rm d} V_{\rm closed} = \int n_e ^2 {\rm d} V_{\rm closed} ,
\end{equation}
where $n_e$ and $n_i$ are the electron and ion number densities, respectively. The integration above is performed in the region of closed field lines (with a volume $V_{\rm closed}$), where the temperature is $\sim 1.5 - 2 \times 10^6$~K. The values obtained are presented in Table~\ref{table}. We found that ${\rm EM} \sim 10^{50.6}~{\rm cm}^{-3}$, which is consistent with observations of \citet{2011A&A...527A.144M}, who found that the emission measure distribution peaks at  $\sim 10^{51}~{\rm cm}^{-3}$. This suggests that our choice for the coronal base density is representative of $\tau$~Boo, implicating that this star might indeed have a denser wind than that of the Sun ($\mdot \approx 135~\mdot_\odot$ according to our models).

Higher $\mdot$ are also predicted/estimated by other authors. In a recent paper, \citet{2011ApJ...741...54C} developed a model that predicts mass-loss rates of cools stars directly from stellar parameters. As $\tau$~Boo is within the range covered by the Cranmer \& Saar scaling, we used their provided IDL routine to compute $\mdot$ predicted by their models. Using the stellar parameters shown in Table~\ref{table_par}, the metallicity (${\rm [Fe/H]}=0.23$) and stellar luminosity ($\log{(L_\star/L_\odot)}=0.481$) provided by \citet{2005ApJS..159..141V}, the scaling relations developed by \citet{2011ApJ...741...54C} predict that $\tau$~Boo should have $\mdot \approx 330 ~\mdot_\odot \approx 6.6 \times 10^{-12}~\msano$. Other estimates were derived by \citet[][$\mdot\approx 83.4~\mdot_\odot \approx 1.67\times 10^{-12}~\msano$]{2005MNRAS.356.1053S} and  \citet[][$\mdot\approx 198.5~\mdot_\odot \approx 3.97 \times 10^{-12}~\msano$]{2010A&A...522A..13R}, who adopted the empirically derived relation between $\mdot$ and the X-ray flux from \citet{2002ApJ...574..412W, 2005ApJ...628L.143W}. These predictions/estimates are consistent with our results, namely that the wind mass-loss rate of $\tau$~Boo should be significantly higher than the solar value. 
 
The wind outflowing along magnetic field lines carries away stellar angular momentum, therefore exerting a braking torque in the star. However, it is known that F-type stars are not very efficient in losing angular momentum. This can be seen, for instance, in the open cluster NGC~6811 which, at an age of 1 Gyr \citep{2011ApJ...733L...9M}, still presents F-type stars that are rapidly rotating, while redder stars (main-sequence G stars) have spun down more considerably. To examine the stellar magnetic braking, we evaluate the angular momentum loss rate carried by the wind of $\tau$~Boo as \citep{1970MNRAS.149..197M}. 
\begin{equation}\label{eq.jdot}
\dot{J} =  \oint_{S_A} \left( p + \frac{B^2}{8\pi} \right) ({\bf r} \times {\bf \hat{n}})_z + \rho {\bf V} \cdot {\bf \hat{n}} \left[ {\bf r} \times (\boldsymbol\Omega \times {\bf r})\right]_z {\rm d} { S_A} ,
\end{equation} 
where ${\bf V} = {\bf u} - \boldsymbol\Omega\times {\bf r}$ is the velocity vector in the frame rotating with angular velocity $\boldsymbol\Omega$, $S_A$ is the Alfv\'en surface, and ${\bf \hat{n}}$ is the normal unit vector to the Alfv\'en surface. We obtained that $\dot{J}$ varies during the observed phases of the cycle, ranging from $1.1$ to $2.2 \times 10^{32}$~erg (Table~\ref{table}). 

We also estimate the time-scale for rotational braking as $\tau = {J}/{\dot{J}}$, where $J$ is the angular momentum of the star. If we assume a spherical star with a uniform density, then $J = 0.4 M_* R_*^2 \Omega_*$ and the time-scale is 
\begin{equation}
\tau \simeq \frac{9 \times 10^{33}}{\dot{J}} \left( \frac{M_*}{M_\odot}\right) \left( \frac{1~{\rm d}}{P_{\rm rot}} \right) \left(  \frac{R_*}{R_\odot}\right)^2 ~{\rm Gyr} 
\end{equation} 
\citep{2011MNRAS.412..351V}. Our results are also shown in Table~\ref{table}. For $\tau$~Boo, spin-down times (spanning from $39$ to $78$~Gyr) are an order of magnitude larger than its age (about $2.4$~Gyr, \citealt{2005A&A...443..609S}), suggesting that, if the stellar wind is the only contributor of redistribution of angular momentum of the star, $\tau$~Boo should maintain its relatively high rotation rate during its main-sequence lifetime. 

The reconstructed maps used here do not have error bars. Therefore, we can not assess the error that is propagated in our simulations due to associated uncertainties from the observations. We evaluate the accuracy in our simulations by calculating the largest variations found across the simulation domain (essentially due to changes in grid resolution). The accuracy varies between cases; in the worst scenario, an accuracy of $1.2$ per cent is found for $\mdot$ and of $16$ per cent for $\jdot$.

Table~\ref{table} also shows the amount of (unsigned) flux in the open magnetic field lines 
\begin{equation}\label{eq.piopen}
\Pi_{\rm open} = \oint_S { |B_r|} {\rm d} S, 
\end{equation}
along which the stellar wind is channelled. Note that the integral in equation~(\ref{eq.piopen}) is performed over a spherical surface $S$. The open flux is calculated sufficiently far from the star ($\gtrsim 10~R_\star$). The fraction of open flux, defined as $f_{\rm open} = \Pi_{\rm open}/\Pi_0$, is also presented in Table~\ref{table}.

A closer inspection of Table~\ref{table} shows that for $\tau$~Boo, mass-loss rates do not vary significantly during the observed epochs of the stellar cycle (variation is at most $3$~per cent), while the angular momentum-loss rate varies by a factor of about $2$.  Similarly to the mass-loss rates, the emission measure shows a negligible variation during the observed epochs of the cycle ($\lesssim 3$ per cent). 

The fact that $\tau$~Boo's calculated emission measure does not vary during the cycle suggests that this star should not present significant variations in its quiescent X-ray emission during its cycle and a magnetic cycle of $\tau$~Boo may not be detected by X-ray observations. Indeed, a recent work points in this direction \citep{2012AN....333...26P}. 

As already mentioned in Section~\ref{sec.maps}, given the similarities of the surface magnetic distributions of $\tau$~Boo derived in June-2006 and July-2008, these maps appear to describe similar cycle phases at distinct magnetic cycles. Because the properties of the stellar wind depend on the particular characteristic of the magnetic field, we expect the stellar wind at these two distinct epochs to be similar. As can be seen in Table~\ref{table}, this is indeed what is found.

\begin{figure*}
\includegraphics[width=84mm]{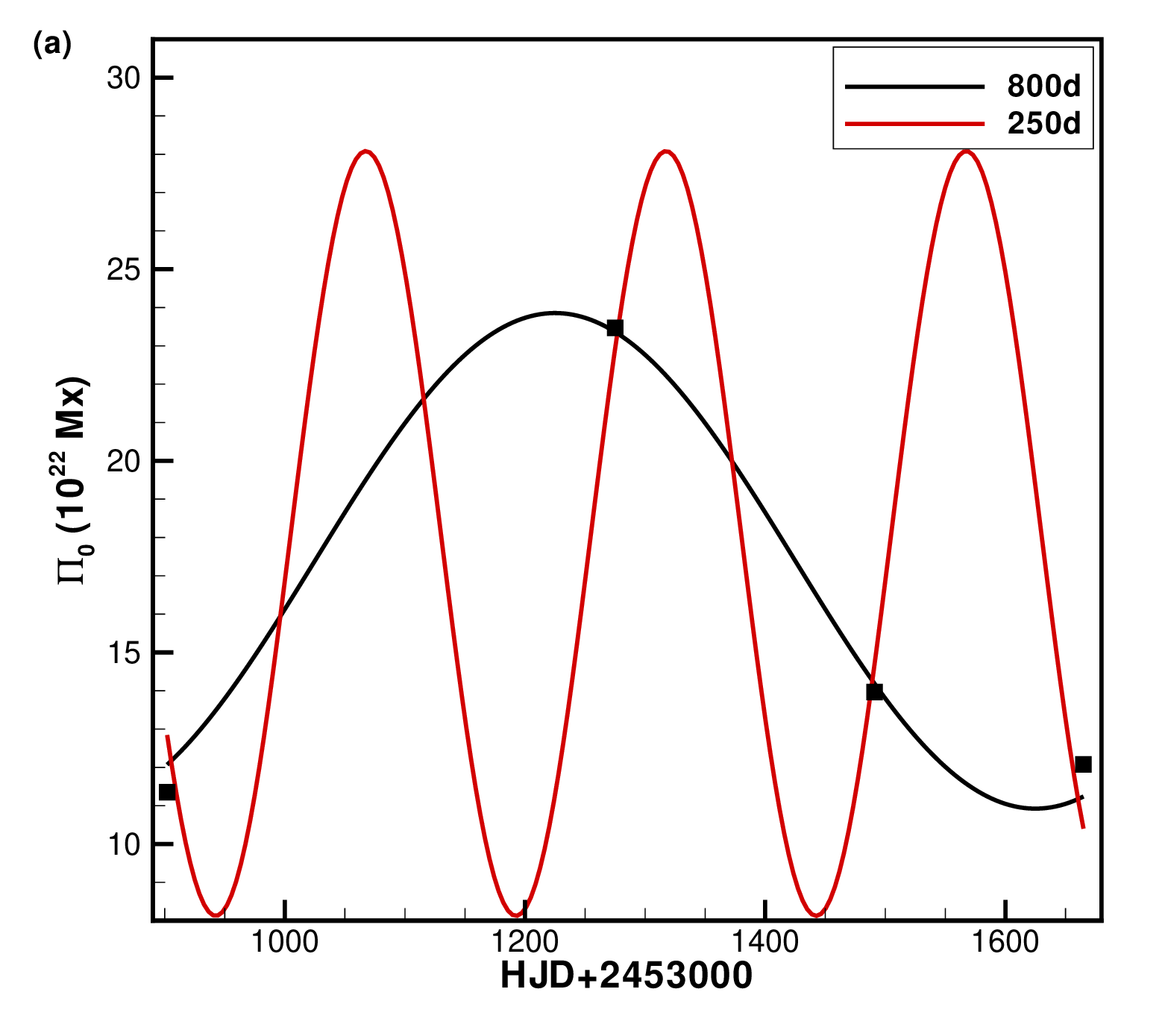}
\includegraphics[width=84mm]{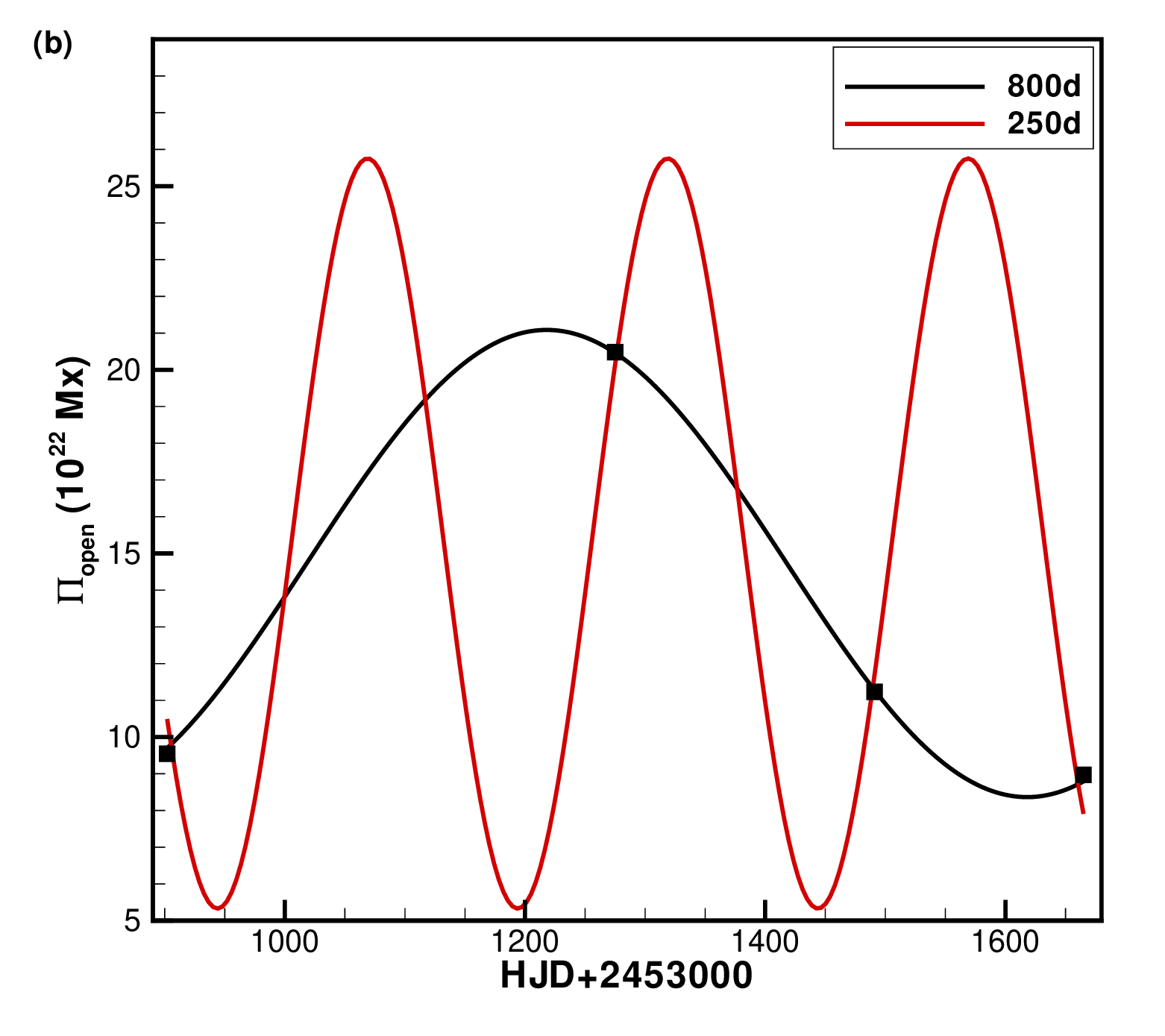}\\
\includegraphics[width=84mm]{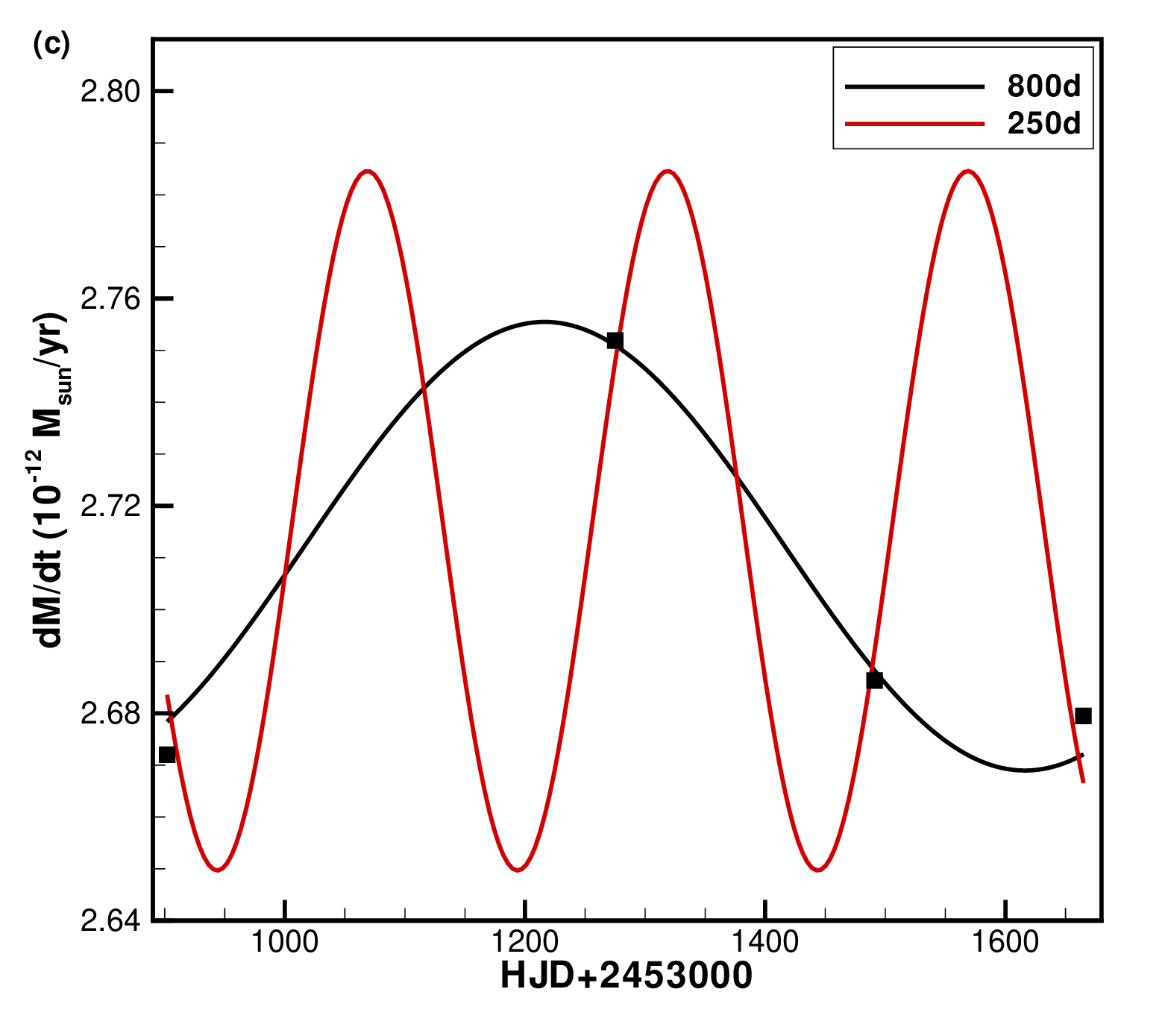}
\includegraphics[width=84mm]{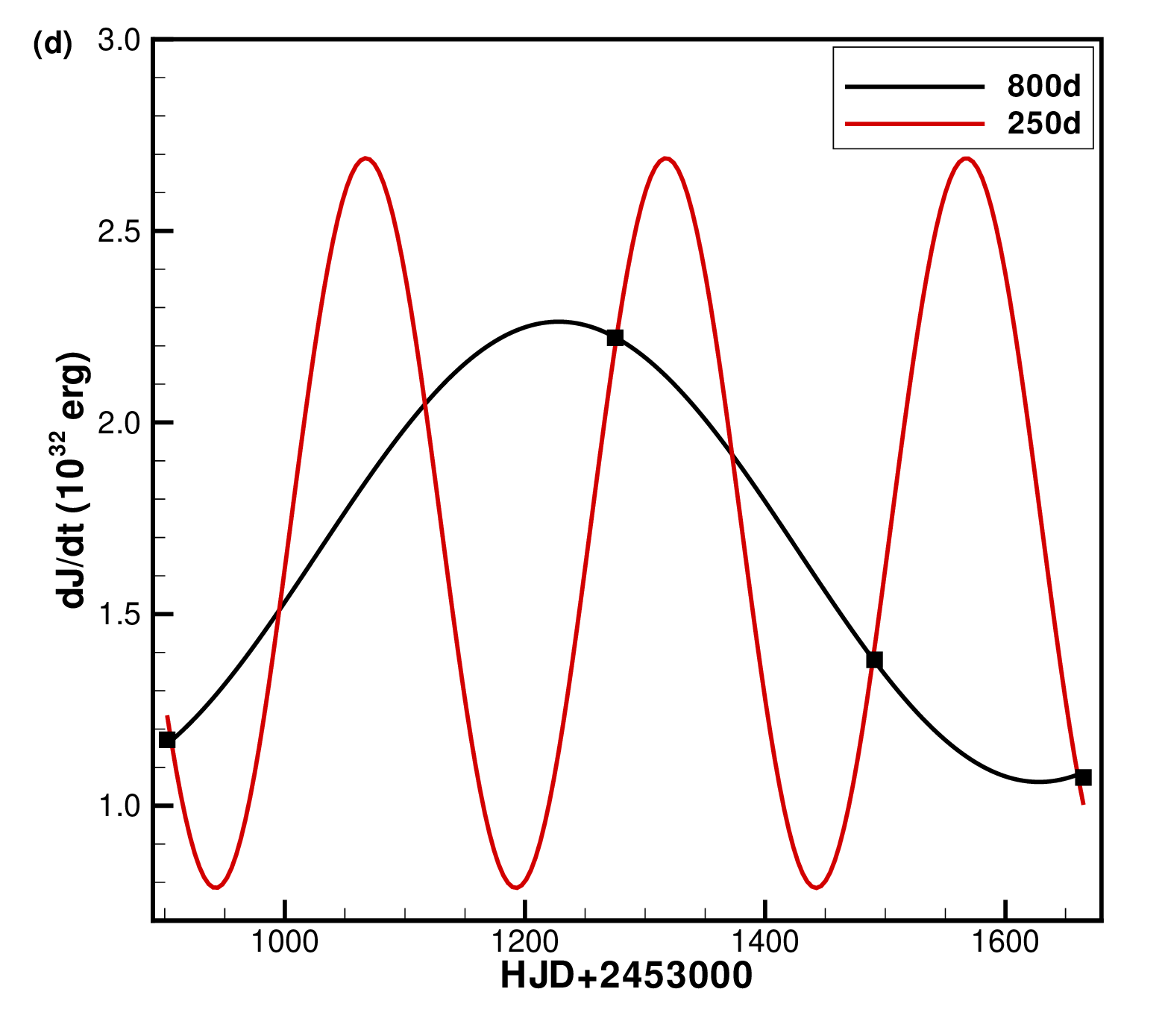}
\caption{Temporal evolution of selected quantities evaluated at four different epochs of the stellar magnetic cycle (black squares, see Table~\ref{table}). Sinusoidal fits to these quantities are shown for a cycle period of 800~d (black solid line) and for 250~d (red solid line). The quantities shown are: (a) surface flux, (b) open flux, (c) mass-loss rates and (d) angular momentum-loss rates. \label{fig.fits}}
\end{figure*}

To investigate the variation during the cycle, we present in Figure~\ref{fig.fits} sinusoidal fits to both the observed data (Fig.~\ref{fig.fits}a) and the results of the wind modelling shown in Table~\ref{table} (Fig.~\ref{fig.fits}b -- d). 
For comparison purposes, using the two cycle periods obtained by \citet{2009MNRAS.398.1383F}, two fits were done for each panel: the black line adopts a cycle period of 800~d, while the red line one of 250~d. Fig.~\ref{fig.fits}a, which shows the variation of the unsigned observed surface flux, illustrates the cyclic nature of the large-scale magnetic field of $\tau$~Boo. The same behaviour is seen in the plot of the open magnetic flux (Fig.~\ref{fig.fits}b). The remaining panels of Figure~\ref{fig.fits} illustrate the evolution of $\mdot$ and $\jdot$ through the stellar magnetic cycle, showing that the stellar wind also behaves in a cyclic way. Note however the very small amplitudes in the variation of $\mdot$.

\section{Results: Radio Emission from Wind-Planet Interaction}\label{sec.radio}
In this section, we present an estimate of radio flux that should arise from the interaction of $\tau$~Boo b's magnetic field with the stellar wind. For that, we use the results of our models, presented in Section~\ref{sec.results_wind}. We note that the lack of knowledge of some properties of the planet, such as its radius or its magnetic field intensity, leads us to adopt some (reasonable) hypotheses. These hypotheses are clearly stated in the following paragraphs. Section~\ref{sec.conc} presents a discussion about how different assumptions would change our results.

The first unknown quantity in our calculations is the orbital inclination of $\tau$~Boo b. The position of the planet is required because the characteristics of the host star wind are three-dimensional in nature. Therefore, the characteristics of the ambient medium surrounding the planet, with which the planet will inevitably interact, depend on the planet's position (orbital distance, longitude and colatitude). It is worth mentioning that, because of the large differential rotation of $\tau$~Boo (the equator rotates with a period of 3~d, while near the poles at 3.9~days), at the colatitude of $\sim 45^{\rm o}$, the planetary orbital period and the stellar surface rotation are similar. 
The orbital ephemeris used in the stellar surface magnetic maps of \citet{2007MNRAS.374L..42C,2008MNRAS.385.1179D, 2009MNRAS.398.1383F} places the conjunction of the planet at the stellar rotation phase $0.0$, which constrain the meridional plane where the planet should be located: in the configuration shown in Figure~\ref{fig.Blines}, this is at the $xz$-plane (more precisely in the region of $x<0$). The orbital radius of the planet has been determined to be $0.0462~{\rm au}$ \citep{butler}, which is about $6.8 ~R_\star$. The orbital inclination of the planet, if known, would constrain at which colatitude the planet is orbiting.  Because of this unknown, the next calculations are computed for a range of colatitudes $\theta$, where $\theta=0^{\rm o}$ is at the rotation pole of the star (in the visible hemisphere) and $\theta=90^{\rm o}$, the equatorial plane of the star. 

\begin{figure}
\includegraphics[width=84mm]{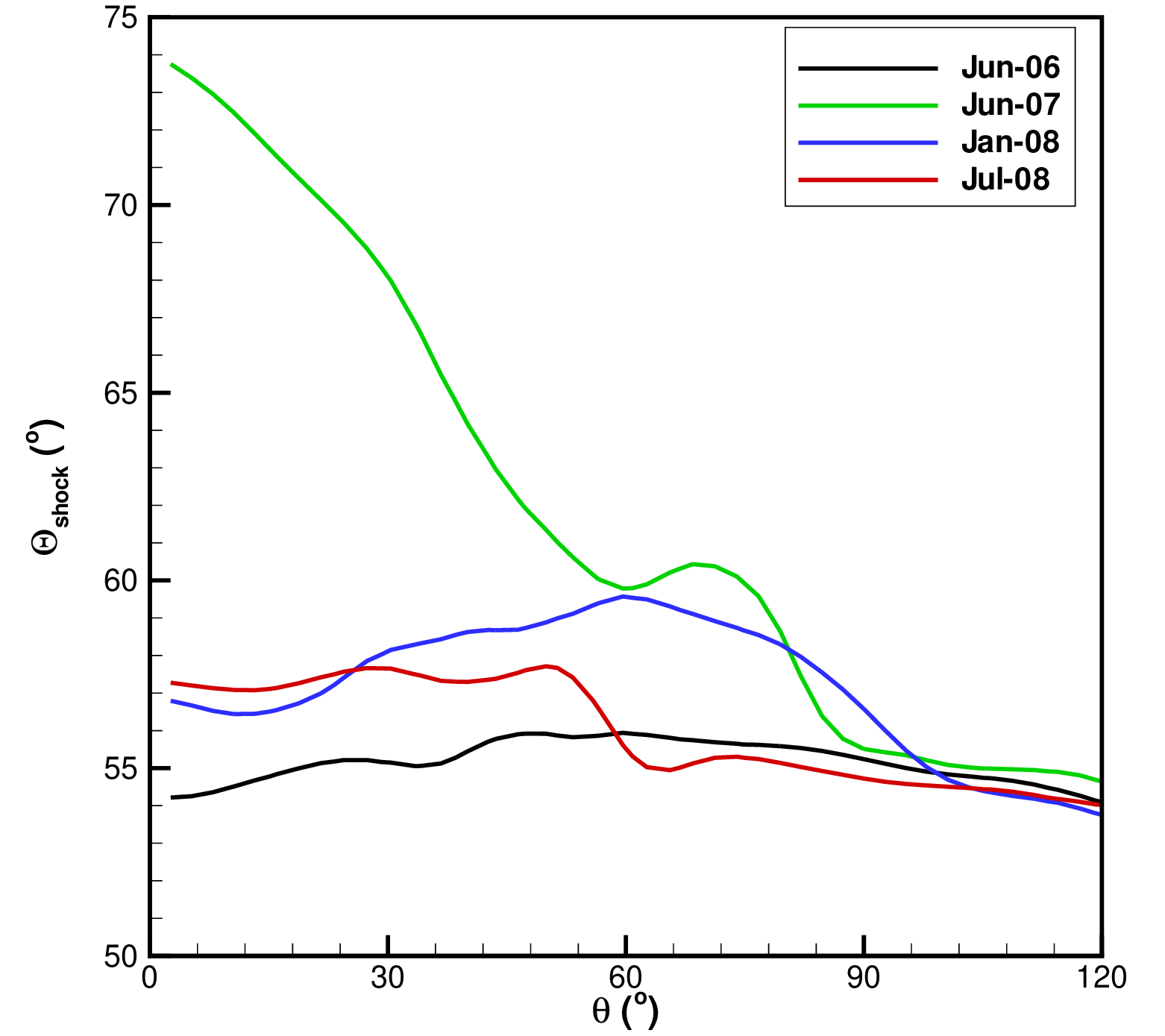}
\caption{Variation of the shock angle $\Theta_{\rm shock}$ for the observed epochs of the stellar cycle as a function of colatitude $\theta$ of the planet orbit. \label{fig.theta_shock}}
\end{figure}

As the planet orbits around its host star, it interacts with the stellar wind, which, in the case of $\tau$~Boo, varies during the stellar magnetic cycle, as a response to variations in the stellar magnetic field (Section~\ref{sec.results_wind}). At the orbital distance of $\tau$~Boo b, the interaction between the planet and the stellar wind takes place at super-magnetosonic velocities, ensuing the formation of a bow-shock around the planet. We define the angle $\Theta_{\rm shock}$ that the shock normal makes to the relative azimuthal velocity of the planet as 
\begin{equation}\label{eq.angle-wind}
\Theta_{\rm shock} =  \arctan{\left(\frac{u_r}{v_{K}- u_{\varphi} }\right) } .
\end{equation}
where we assume the planet to be at a circular Keplerian orbit ($v_K = (G M_\star/r_{\rm orb})^{1/2}$). When the shock normal points to the host star, $\Theta_{\rm shock}=90^{\rm o}$ and the shock is a dayside shock \citep{2010ApJ...722L.168V}, similar to the one that surrounds the Earth's magnetosphere \citep{1998ISSIR...1..249S}. On the other hand, for a shock normal pointing ahead of the planetary orbit, $\Theta_{\rm shock}=0^{\rm o}$ and the shock is an ahead shock \citep{2010ApJ...722L.168V}. Figure~\ref{fig.theta_shock} shows how the angle $\Theta_{\rm shock}$ varies through the observed phases of the stellar cycle for a range of colatitudes $\theta$. We see that the shock formed around $\tau$~Boo b's magnetosphere forms at angles $52^{\rm o} \lesssim \Theta_{\rm shock} \lesssim 74^{\rm o}$.

\begin{figure*}
\includegraphics[width=84mm]{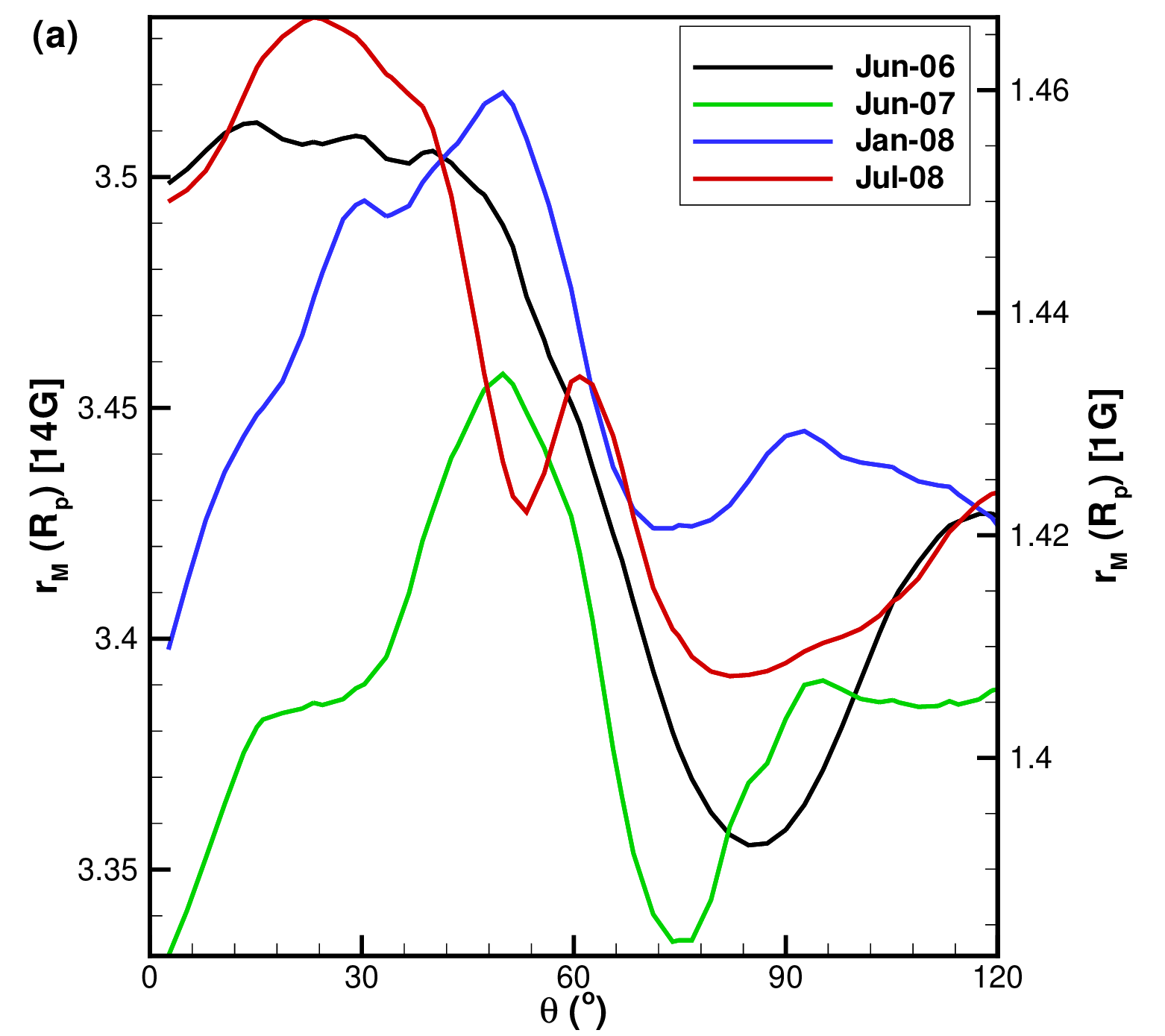}
\includegraphics[width=84mm]{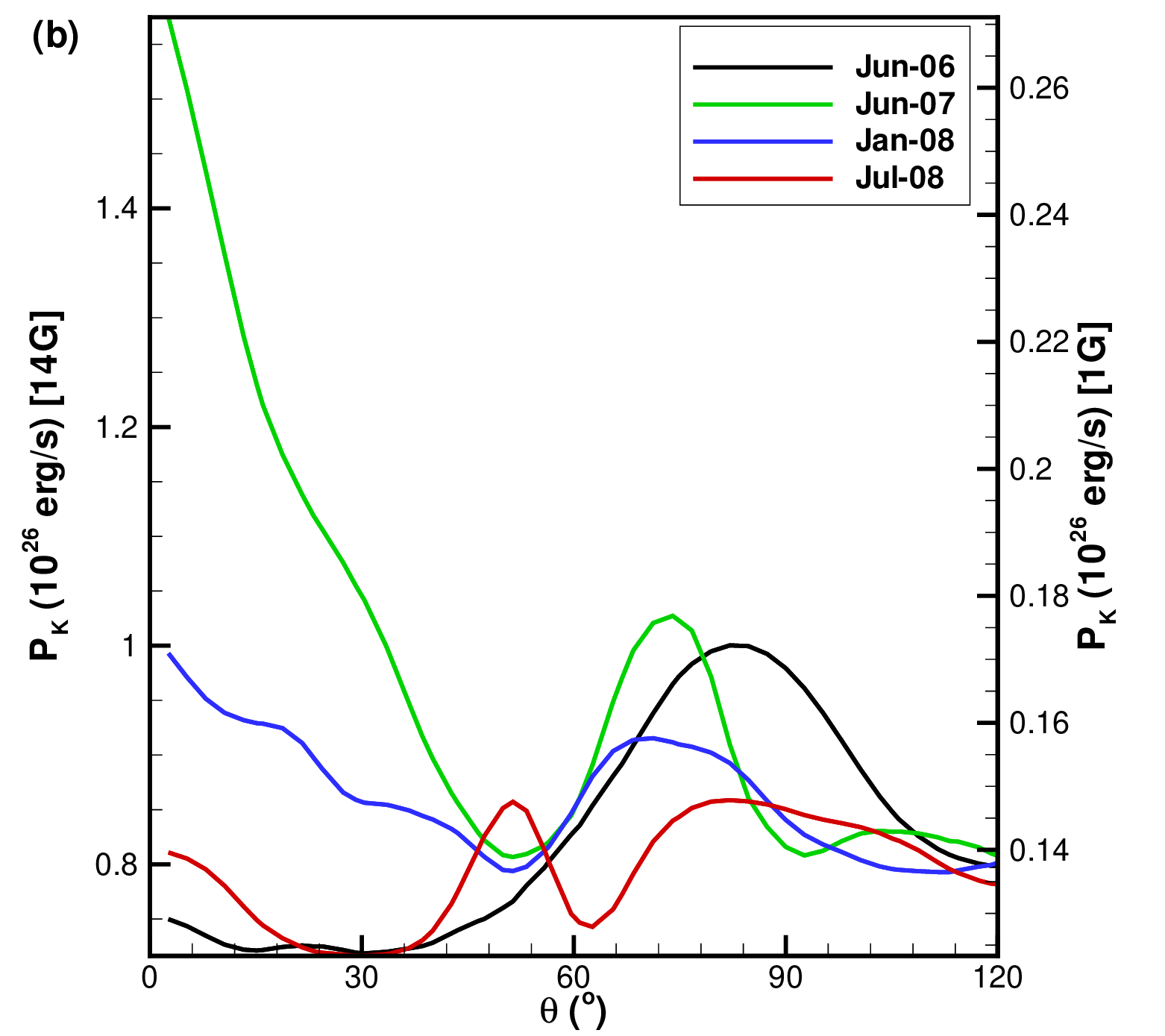}\\
\includegraphics[width=84mm]{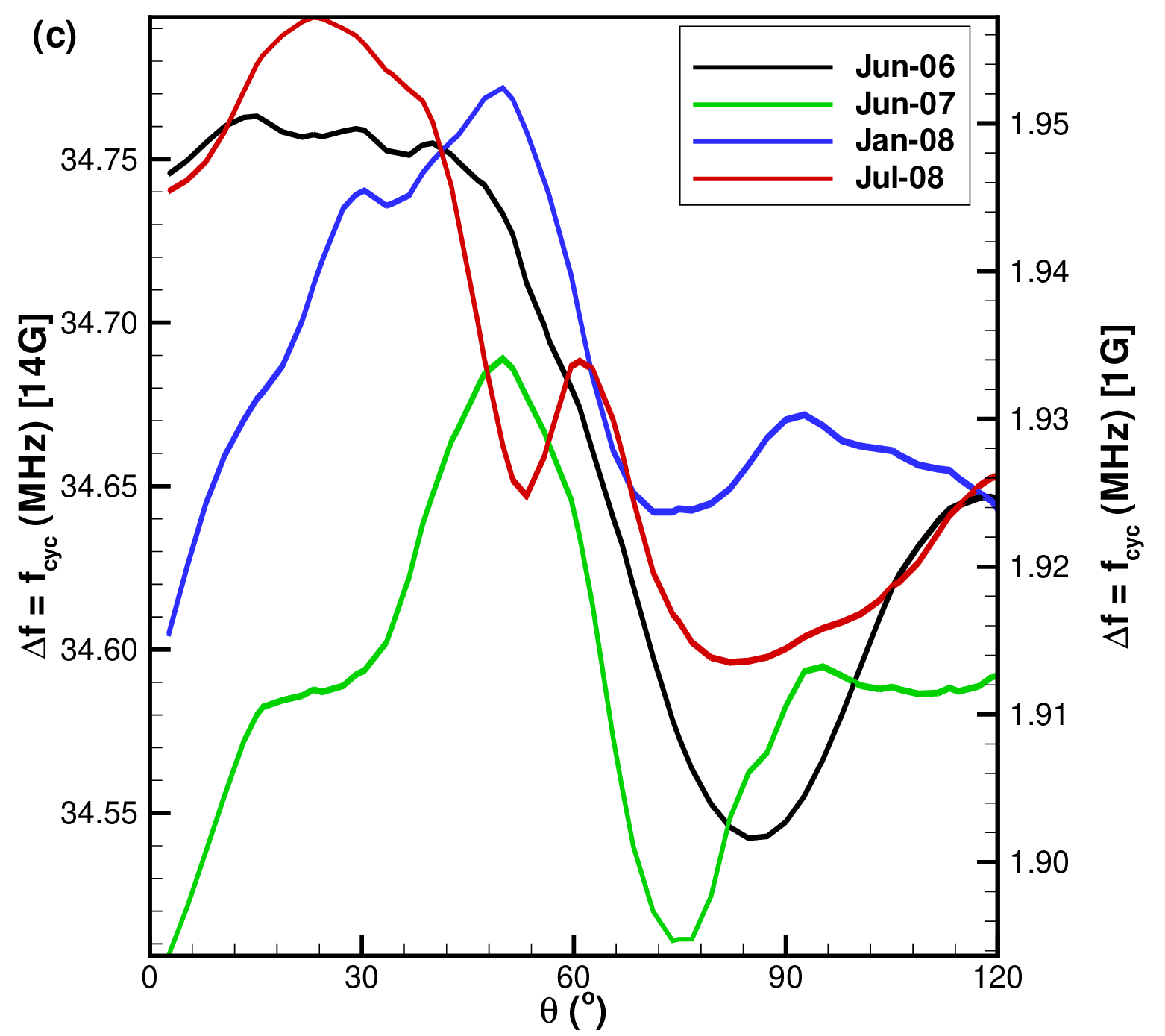}
\includegraphics[width=84mm]{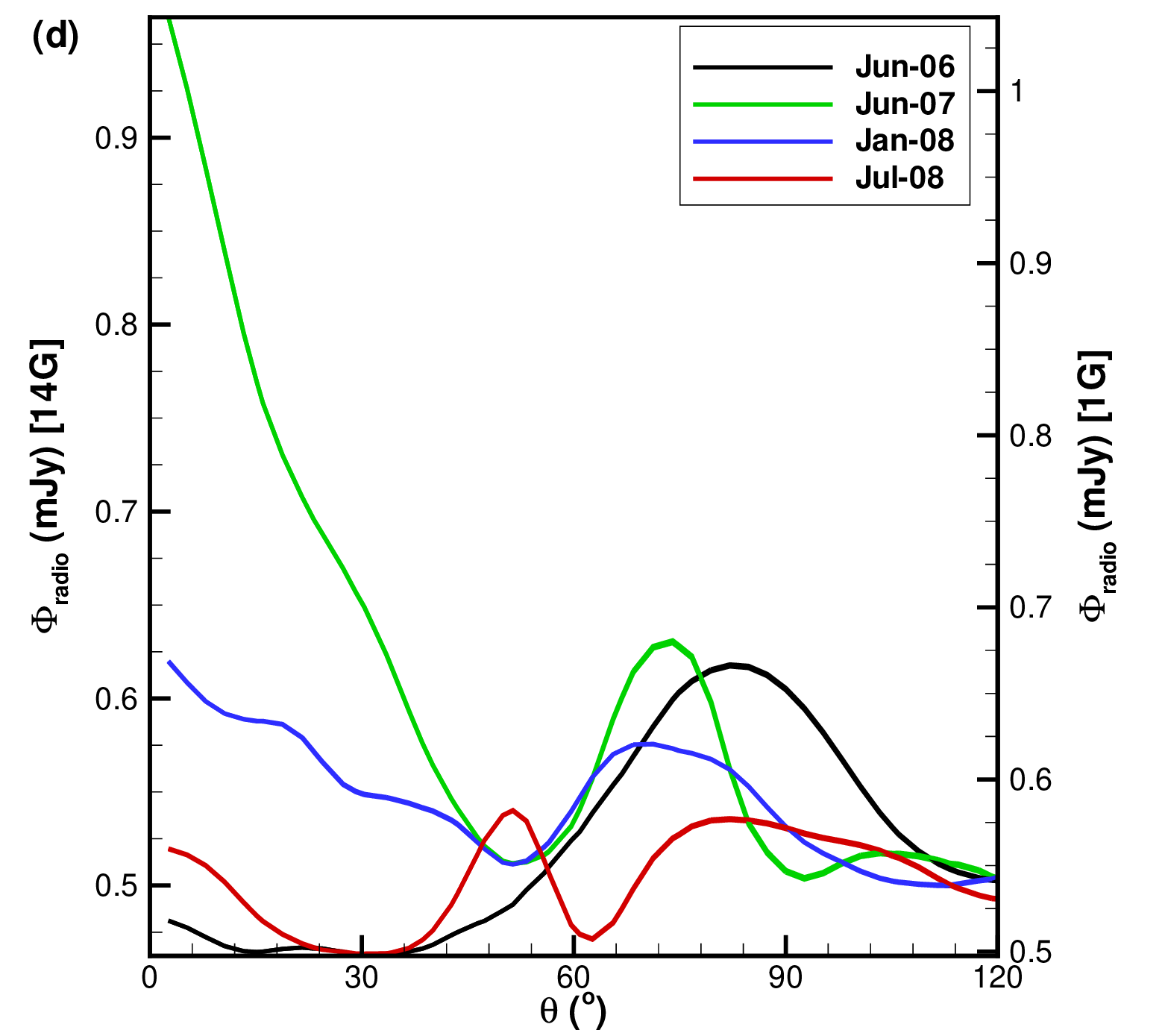}%
\caption{(a) Magnetospheric radius of $\tau$~Boo b as a function of colatitude $\theta$, which is related to the planet's unknown orbital inclination (see text). The planet is either assumed to have a magnetic field intensity similar to the Earth (right axis) or similar to Jupiter's (left axis). (b) Wind kinetic power impacting in the planet. (c) Frequency of the radio emission. (d) Radio flux estimated for $\tau$~Boo b. \label{fig.radio}}
\end{figure*}

The magnetosphere of the planet deflects the stellar wind around it, forming a cavity in the wind. The extent of the planetary magnetosphere $r_M$ can be determined by static pressure balance \citep[e.g.,][]{2011MNRAS.414.1573V}
\begin{equation}\label{eq.equilibrium}
{\rho \Delta u^2} + \frac{B^2}{8\pi} + p= \frac{B_{p}^2(r_M)}{8\pi} + p_{p} .
\end{equation}
where $|\Delta {\bf u}| = |{\bf u} - {\bf v}_{K}|$ is the relative velocity between the wind and the planet, and $\rho$, $p$ and $B$ are the local density, pressure and magnetic field intensity at the ambient medium surrounding the planet. $B_{p}(r_M)$ is the intensity of the planet's magnetic field at the nose of the magnetopause. In Equation~(\ref{eq.equilibrium}), we neglect compression of the planetary magnetic field (pile-up). In our next calculations, we neglect the planet thermal pressure $p_p$. We assume that the planetary magnetic field is dipolar, such that $B_{p}(r_M) = B_{p, eq} (R_p/R)^3$, where $R_p$ is the planetary radius, $R$ is the radial coordinate centred at the planet, $B_{\rm p, eq}=B_p/2$ is the magnetic field intensity  evaluated  at the equator of the planet and $B_{p}$  at its pole. Assuming the dipole is aligned with the planetary orbital spin axis, the magnetospheric radius (i.e., where $R=r_M$) is given by
\begin{equation}\label{eq.rM}
\frac{r_M}{R_p}=\left[ \frac{(B_p/2)^2}{{8\pi(\rho \Delta u^2}  + p) + B^2} \right]^{1/6} .
\end{equation}
Figure~\ref{fig.radio}a shows how the magnetospheric radius of $\tau$~Boo b varies through the observed epochs of the stellar magnetic cycle. The magnetic field of the planet is a quantity that has not yet been directly observed for extrasolar planets. 
If confirmed, the technique proposed by \citet{2010ApJ...722L.168V}, based on near-UV transit observations, should provide a useful tool in determining planetary magnetic field intensities for transiting systems. In their estimates for WASP-12b (the only case for which near-UV data is available so far), they placed an upper limit on the intensity of the planetary magnetic field of about the same order of magnitude as Jupiter's magnetic field. Dynamo models of \citet{2004ApJ...609L..87S} suggest that close-in giant planets should present a magnetic field intensity similar to the Earth's.  
To accommodate these two suggestions, we assume the planet to have a magnetic field intensity similar to that of the Earth ($B_p=1$~G, right axis in Figure~\ref{fig.radio}a) and Jupiter ($B_p=14$~G, left axis). For a terrestrial magnetic field intensity, the planet's magnetospheric size is quite small ($1.36 < {r_M}/{R_p}< 1.46$), while it is a few times larger for a jovian magnetic field intensity ($3.2 < {r_M}/{R_p}< 3.6$). Because the characteristics of the stellar wind is the same for both cases (left side of Equation~\ref{eq.equilibrium}), the smallest magnetospheric radius for the case with $B_p=1$~G is due to the smallest magnetic moment assumed. We note that the former case is quite similar to the size of the magnetosphere of Mercury \citep[$\sim 1.3 - 1.9$ Mercury radii][]{1988merc.book..514R}. For comparison, the radius of the Earth's magnetosphere is $\sim 10-15~R_{\rm Earth}$ \citep{2002P&SS...50..549S}. Furthermore, we note that the magnetospheric size variations through the cycle are at most of $\sim 5$~per cent depending on the colatitude $\theta$  (related to the orbital inclination).

In the solar system, the planetary radio power $P_{\rm radio}$ is related to the impacting solar wind kinetic  and magnetic  powers ($P_k$ and $P_B$, respectively). Kinetic-to-radio efficiency is $P_k/P_{\rm radio} = 10^{-5}$ and magnetic-to-radio efficiency is $P_B/P_{\rm radio} = 2\times 10^{-3}$ \citep{2007P&SS...55..598Z}. We assume that the same efficiency ratios will hold in the $\tau$~Boo planetary system.

The kinetic power of the impacting wind on the planet is approximated as the ram pressure of the wind ($\rho (\Delta u)^2$) impacting in the planet, whose area is $\pi r_M^2$, at a (relative) velocity $\Delta u$ 
\begin{equation}\label{eq.pK}
P_k \simeq \rho (\Delta u)^3 \pi r_M^2 .
\end{equation}
By using the results of our simulations in the previous equation, we are thus able to compute the impacting kinetic power of the wind. Note that our models can only obtain the magnetospheric size of the planet relative to its radius (Fig.~\ref{fig.radio}a). To convert the relative magnetospheric size to a physical value we need information on the unknown radius of the planet $R_p$ (note that $\tau$~Boo b is not transiting its host star). Following \citet{2004ApJ...609L..87S} and \citet{2007P&SS...55..618G}, we adopt $R_p=1.3 R_{\rm jup}$. Figure~\ref{fig.radio}b shows how $P_k$ varies through the observed epochs of the stellar magnetic cycle. With the assumed kinetic-to-radio efficiency as that observed for the solar system planets ($P_{\rm radio}\sim 10^{-5} P_k$), we conclude that radio power emitted by $\tau$~Boo b should be $P_{\rm radio} \simeq 0.7 - 1.9 \times 10^{14}$~W for $B_p=14$~G and $P_{\rm radio} \simeq 0.12 - 0.33 \times 10^{14}$~W for $B_p=1$~G.

Because the frequency of the radio emission is related to the cyclotron frequency, the magnitude of the planet's magnetic field is required to estimate the bandwidth $\Delta f$ of the radio emission. 
In Jupiter, decametric emission is thought to arise in a ring surrounding the auroral region wherein the planetary magnetic field lines are open. The aperture of the auroral ring can be related to the size of the planet's magnetosphere $r_M$ as
\begin{equation}\label{eq.alpha}
\alpha_0 = \arcsin \left[ \left( \frac{R_p}{r_M} \right)^{1/2} \right]
\end{equation}
where $\alpha_0$ is the colatitude of this auroral ring \citep[more details in][]{2011MNRAS.414.1573V}. The planetary magnetic field at colatitude $\alpha_0$ is
\begin{equation}\label{eq.B_alpha0}
B(\alpha_0) = \frac{B_p}{2} (1+3\cos^2 \alpha_0)^{1/2}.
\end{equation}
We assume that the emission bandwidth $\Delta f$ is approximately the cyclotron frequency \citep{2007P&SS...55..618G}, where
\begin{equation}\label{eq.fcyc}
\Delta f = f_{\rm cyc}= 2.8 \left( \frac{B(\alpha_0)}{1~{\rm G}}\right) ~{\rm MHz} .
\end{equation}
Figure~\ref{fig.radio}c shows the predicted emission bandwidth from $\tau$~Boo b assuming $B_p=14$~G ($\Delta f \simeq 34 $~MHz) and assuming $B_p=1$~G  ($\Delta f \simeq 2$~MHz). As the ionospheric cut-off is at frequencies $\lesssim 10$~MHz, we note that if $\tau$~Boo b has a magnetic field similar to that of the Earth, we would not be able to detect any planetary emission from the ground. Using Equation~(\ref{eq.fcyc}), we roughly estimate the minimum planetary magnetic field intensity required for the radio frequency to lie above the cut-off value of $10~$MHz to be $B_{\rm min} \sim 4$~G. We note that, for $B_p= 14$~G, the predicted $\Delta f \simeq 34 $~MHz lies in the observable range of LOFAR. 

The radio flux is related to the radio power as
\begin{equation}\label{eq.radio}
\phi_{\rm radio} = \frac{P_{\rm radio}}{d^2 \omega \Delta f}
\end{equation}
where $d=15.6$~pc is the distance to the system and $\omega= 2\times 2 \pi (1 - \cos \alpha_0)$ is the solid angle of the emission (the factor of two was included in order to account for emission coming from both Northern and Southern auroral rings). The radio flux is presented in Figure~\ref{fig.radio}d which shows that $\phi_{\rm radio} \simeq 0.5-1$~mJy.\footnote{We demonstrate below that $\phi_{\rm radio}$ weakly depends on the planetary magnetic field $B_p$ for the parameters adopted. First, we note that $\alpha_0$ correlates to $B_p$ through Equation~(\ref{eq.rM}) as: $ \sin \alpha_0 = (B_p^2/\mathcal{C})^{-1/12}$, where for the stellar winds simulated here, $\mathcal{C} \simeq 0.10-0.16$. This results in $\alpha_0 \simeq 32^{\rm o}$ for $B_p=14~$G and $\alpha_0 \simeq 57^{\rm o}$ for $B_p=1~$G. In addition, it is easy to see from Equations~(\ref{eq.pK}) and (\ref{eq.alpha}) that $P_{\rm radio}\propto P_k \propto r_M^2 \propto \sin^{-4} \alpha_0$. Substitution of the previous results, $\omega \propto (1 - \cos \alpha_0)$,  Equations~(\ref{eq.B_alpha0}) and (\ref{eq.fcyc}) in (\ref{eq.radio}) reveals that:  $\phi_{\rm radio}\propto \sin^{-4} \alpha_0/[B_p(1 - \cos \alpha_0) (1+3\cos^2 \alpha_0)^{1/2} ]  \propto (1 + \cos \alpha_0)  (1+3\cos^2 \alpha_0)^{-1/2}$, which has a weak dependence in $\alpha_0$.  For instance, for  $32^{\rm o} \lesssim \alpha_0 \lesssim 57^{\rm o}$, the function $(1 + \cos \alpha_0)  (1+3\cos^2 \alpha_0)^{-1/2}$ is in the range $[1.04,1.12]$.}

There are in the literature some estimates of radio emission from $\tau$~Boo b. In the most optimistic consideration made by \citet{1999JGR...10414025F}, the median radio flux of $\tau$~Boo b is estimated to be about $2.2$~mJy, at a frequency of $28$~MHz. \citet{2005A&A...437..717G} suggest a radio flux of $4-9$~mJy at a frequency of $7-19$~MHz. Although their assumptions for the stellar wind differ from our adopted model, the results obtained here are comparable to those from \citet{1999JGR...10414025F} and \citet{2005A&A...437..717G}.

\citet{2007ApJ...668.1182L} observed $\tau$~Boo with the VLA at a frequency of $74$~MHz ($4$-m wavelength).  Observations were held at four epochs and radio emission was not detected at any of these epochs above a limit of about 100 to 300 mJy. The estimates presented here are in accordance to these observational findings as, indeed, our estimates predict an order of magnitude smaller flux  at a different frequency. 

We note that, for a radio emission process that is powered by reconnection events between the planetary magnetic field and the stellar coronal magnetic field, certain configurations of the magnetic fields may not favour reconnection. This should be the case, for instance, of an idealised situation where both the planet's and the stellar magnetic field are perfectly aligned and with the same polarity. Therefore, it is possible that, due to the presence of the polarity reversals in the stellar magnetic field, radio emission from $\tau$~Boo b is an intermittent process. 

While we have modelled the stellar wind that flows along the large-scale coronal magnetic field lines, the environment surrounding a star is likely to be much more dynamic, especially for a star such as $\tau$~Boo that is more active than the Sun \citep{2011A&A...527A.144M}. We note that the radio flux estimated here is expected to increase if the planet is hit by powerful ejections of coronal material, e.g., caused by flares or coronal mass ejections.

\section{Discussion and Conclusion }\label{sec.conc}
In this paper, we investigated the variation of the stellar wind in $\tau$~Boo during its cycle by means of three-dimensional numerical simulations. Our simulations adopt observationally derived surface magnetic field maps obtained at four different epochs. Because the stellar wind properties depend on the characteristics of the stellar magnetic field (geometry and intensity), the wind varies during the magnetic cycle of $\tau$~Boo. 

We found that wind mass loss-rate varies little during the observed epochs of the cycle (less than 3 per cent), with a relatively more important variation in angular momentum loss-rates (a factor of 2 during these epochs). The amount of (unsigned) open flux in the magnetic field lines shows a variation of up to a factor of $2.3$ during the epochs studied here. The computed mass loss-rate for $\tau$~Boo is $\mdot \simeq 2.7 \times 10^{-12}~\msano$, two orders of magnitude larger than that of the solar wind. Angular momentum-loss rates vary through the observed epochs and range from $\jdot \simeq 1.1$ to $2.2 \times 10^{32}$ erg, which correspond to characteristic spin-down times $\tau \simeq 39 - 78  ~{\rm Gyr}$, due to the stellar wind alone. 

We also computed the emission measure from the quiescent closed corona, and found that it remains approximately constant through the cycle at a value of ${\rm EM} \simeq 10^{50.6}~{\rm cm}^{-3}$. This suggests that a magnetic cycle of $\tau$~Boo may not be detected by X-ray observations. 

Although several efforts have been made towards detection of auroral radio emission from exoplanets, it has not been detected so far. Radio emission can be pumped by reconnection between the magnetic field lines of the stellar corona and the magnetosphere of the planet. Based on the analogy to the giant planets in the solar system, which shows that radio emission scales with the kinetic and magnetic powers of the incident solar wind, we estimated radio emission from the hot-Jupiter that orbits at $0.0462$~au from $\tau$~Boo. We showed that, for a planet with a magnetic field similar to Jupiter ($B_p\simeq 14$~G), the radio flux is estimated to be about $\phi_{\rm radio} \simeq 0.5-1$~mJy, occurring at an emission bandwidth of $\Delta f \simeq 34$~MHz. Although small, this emission bandwidth lies in the observable range of current instruments, such as LOFAR. However, we note that to observe such a small flux, an instrument with a sensitivity lying on a mJy level is required\footnote{The nominal noise level of LOFAR operating at $30$~MHz, for an exposure time of $1$~h is about 10~mJy. See details in http://www.astron.nl/radio-observatory/astronomers/lofar-imaging-capabilities-sensitivity/sensitivity-lofar-array/sensiti }. The same estimate was done considering the planet has a magnetic field similar to the Earth ($B_p\simeq 1$~G). Although the radio flux does not present a significant difference to what was found for the previous case, the emission bandwidth ($\Delta f \simeq 2$~MHz) falls at a range below the ionospheric cut-off, preventing its possible detection from the ground. In fact, we estimate that, due to the ionospheric cutoff at $\sim 10~$MHz, radio detection  with ground-based observations from planets with $B_p\lesssim 4$~G (Eq.~\ref{eq.fcyc}, with $\Delta f = f_{\rm cyc}$) should not be possible.

We remind the reader that in the estimate of radio emission, the lack of knowledge of some properties of the planet, such as its radius or its magnetic field intensity, and of the efficiency of the radio emission process, led us to make some assumptions, which were clearly stated in Section~\ref{sec.radio}. Although we believe them to be reasonable hypotheses, they may incorporate uncertainties in our calculation, which we discuss next. (1) We assumed that the planet is about 1.3 times the size of Jupiter. The power emitted by the wind (which is converted in radio power) scales as the square of the size of the planet. Therefore, if we had assumed a planet radius of $1.58~R_{\rm Jup}$, the radio flux would increase about 50~per cent the values presented in this paper. (2) A second assumption, maybe the most uncertain one, is the efficiency ratio between the impacting wind power to the emitted radio power. In our estimates, we simply adopted the solar system value of $10^{-5}$, but this is an ad-hoc assumption. Of course, a larger (smaller) efficiency value implies in a larger (smaller) radio flux. (3) We assumed that the emission bandwidth is $\Delta f = f_{\rm cyc}$. Some authors adopt $\Delta f = f_{\rm cyc}/2$ instead, which also could increase the radio flux by a factor of 2 (Eq.~\ref{eq.radio}). (4) A fourth assumption that was implicit in our calculation is that the planet is magnetised, which may not be the case. 

It is difficult to estimate errors involved in the calculated radio flux from $\tau$~Boo b, in especial due to reason (2). Although the detection of small fluxes, such as the ones found in this study, are certainly challenging, modern-day instruments, such as LOFAR, have great potential to detect radio emission from exoplanets. Radio observations of $\tau$~Boo b is, therefore, a valuable exercise.

\section*{Acknowledgements}
AAV acknowledges support from the Royal Astronomical Society through a post-doctoral fellowship. AAV would like to thank J.-M. Griessmeier for useful discussions and providing comments to the manuscript.

\def\aj{{AJ}}                   
\def\araa{{ARA\&A}}             
\def\apj{{ApJ}}                 
\def\apjl{{ApJ}}                
\def\apjs{{ApJS}}               
\def\apss{{Ap\&SS}}             
\def\aap{{A\&A}}                
\def\aapr{{A\&A~Rev.}}          
\def\aaps{{A\&AS}}              
\def\mnras{{MNRAS}}             
\def\pasp{{PASP}}               
\def\solphys{{Sol.~Phys.}}      
\def\sovast{{Soviet~Ast.}}      
\def\ssr{{Space~Sci.~Rev.}}     
\def\nat{{Nature}}              
\def\iaucirc{{IAU~Circ.}}       
\def\planss{{Planetary Space Science}}
\def\jgr{{Journal of Geophysical Research}}   
\def\grl{{Geophysical Research Letters}}

\let\astap=\aap
\let\apjlett=\apjl
\let\apjsupp=\apjs
\let\applopt=\ao

\appendix
\section{Departure from Potential Field}\label{appendix}
 The potential field source surface method  \citep[PFSSM,][]{1969SoPh....9..131A, 2002MNRAS.333..339J} assumes that the magnetic field is everywhere potential, with a surface distribution of $B_r$ derived from observed surface magnetic maps. The greatest advantage of this method is that the coronal magnetic field structure can be computed in a much smaller time-scale than the full MHD solution (for the cases run here, typically seconds versus days). However, a recurrent criticism that the PFSSM faces is that it may not correctly depict the structure of the magnetic field lines, which for instance may not be potential. With the aim of contrasting the results of our MHD modelling of the wind of $\tau$~Boo with the output of the PFSSM, which was used as the initial configuration for the magnetic field lines of our simulations, we compare the energy density of the magnetic field lines as derived by both methods.  

\begin{figure}
\includegraphics[width=84mm]{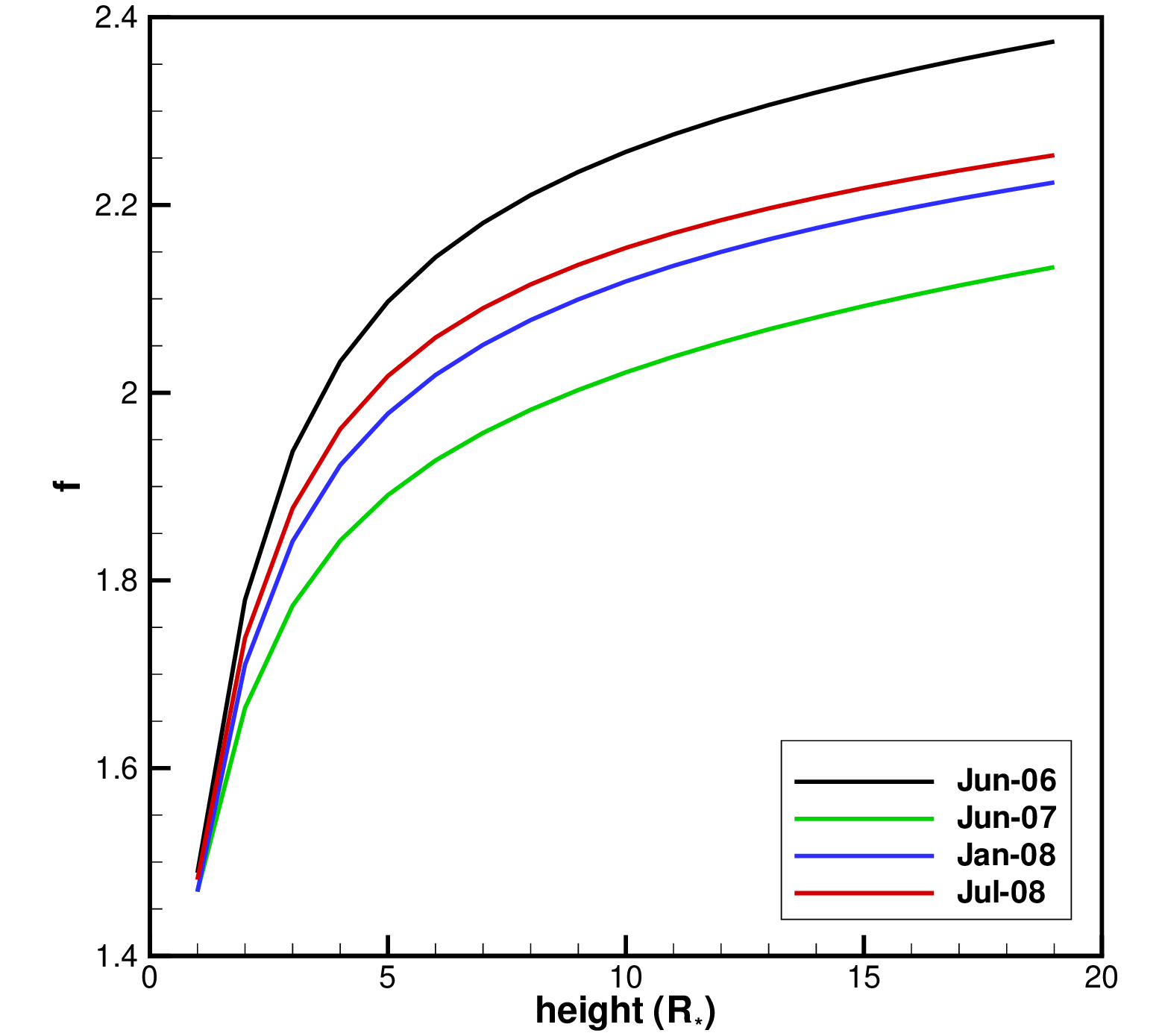}%
\caption{Ratio of the magnetic energy densities of the full MHD solution and the potential field solution (Eq.~\ref{eq.f}) as a function of height. We note that closer to the star, the MHD solution deviates little from the potential field solution, but this deviation becomes more important farther out from the star.\label{fig.Emag}}
\end{figure}

The stored magnetic energy contained in the potential field is in the lowest state, i.e., it is the minimum value of energy that the magnetic field lines can store. In the MHD wind case, excess energy is contained in the magnetic field lines due to stresses imposed by the wind. To quantify the departure of the MHD solution from the potential field solution, we evaluate the stored magnetic energy in each case. Defining the mean magnetic energy as 
\begin{equation}
\langle B^2 \rangle  = \frac{\int_\mathcal{V} B^2 {\rm d} \mathcal{V}}{\int_\mathcal{V} {\rm d}\mathcal{V}},
\end{equation}
where $\mathcal{V}$ is a given spherical volume, we calculated the ratio $f$ between the energy contained in the MHD solution and the one in the PFSSM solution as
\begin{equation}\label{eq.f}
f = \frac{\langle B^2 \rangle_{\rm MHD}}{\langle B^2 \rangle_{\rm PFSSM}}  .
\end{equation}
Figure~\ref{fig.Emag} shows the fraction $f$ as a function of stellar height for the cases we have simulated. To calculate the solution of the PFSSM, we assume that the source surface is located at $4~R_\star$. We find that closer to the star, the MHD solution deviates little from the potential field solution. However, the departure from a potential field becomes more important farther out from the star. Note that at a height of $1~R_*$ above the stellar surface, the stored magnetic energy density in the MHD solution is $\sim 50$ per cent larger than the  magnetic energy density contained in the potential field solution (lowest energy state). At about a height of  $5~R_*$, ${\langle B^2 \rangle_{\rm MHD}}$ is about twice the  value of ${\langle B^2 \rangle_{\rm PFSSM}}$. 

\section{Constraints imposed on the magnetic field reconstruction}\label{sec.assumptions}
As the region of the star with latitudes $\lesssim-40^{\rm o}$ are hidden from the observer, the magnetic field there inevitably depends on assumptions involved in the reconstruction method. The surface magnetic field maps that are shown in Figure~\ref{fig.magnetograms} were obtained without assuming any constraints on the symmetry properties of the field. This results in (unsigned) magnetic field intensities in the unseen hemisphere that are  much smaller and ``smoother'' than the ones found in the visible hemisphere. Here, we investigate the sensitivity of our results with respect to this observational uncertainty. For that, we consider two different constraints on the properties of the magnetic field distribution that could have been applied during the reconstruction, if a physically-motivated reason existed. In this investigation, we consider the observing epoch of July-2008. 

\begin{figure}
\includegraphics[width=84mm]{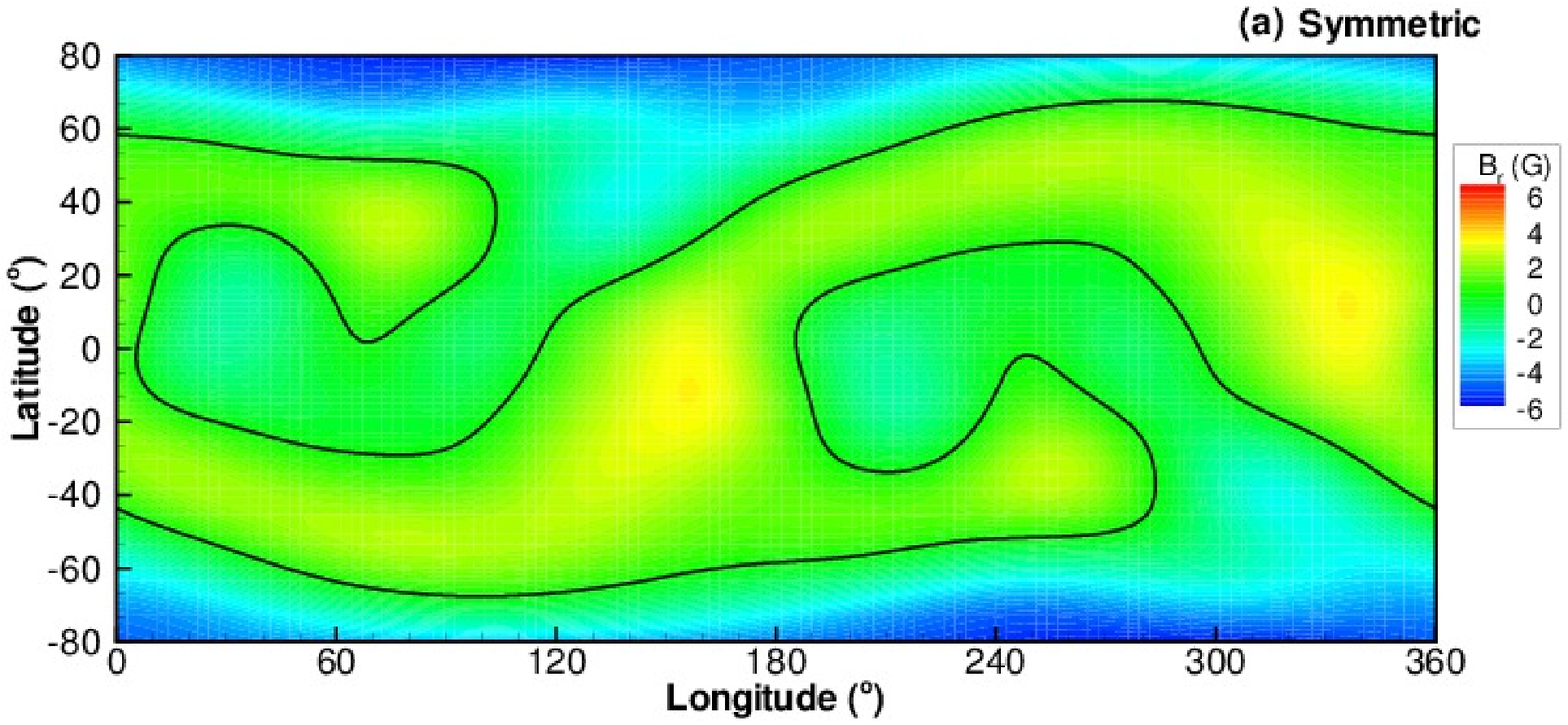}\\
\includegraphics[width=84mm]{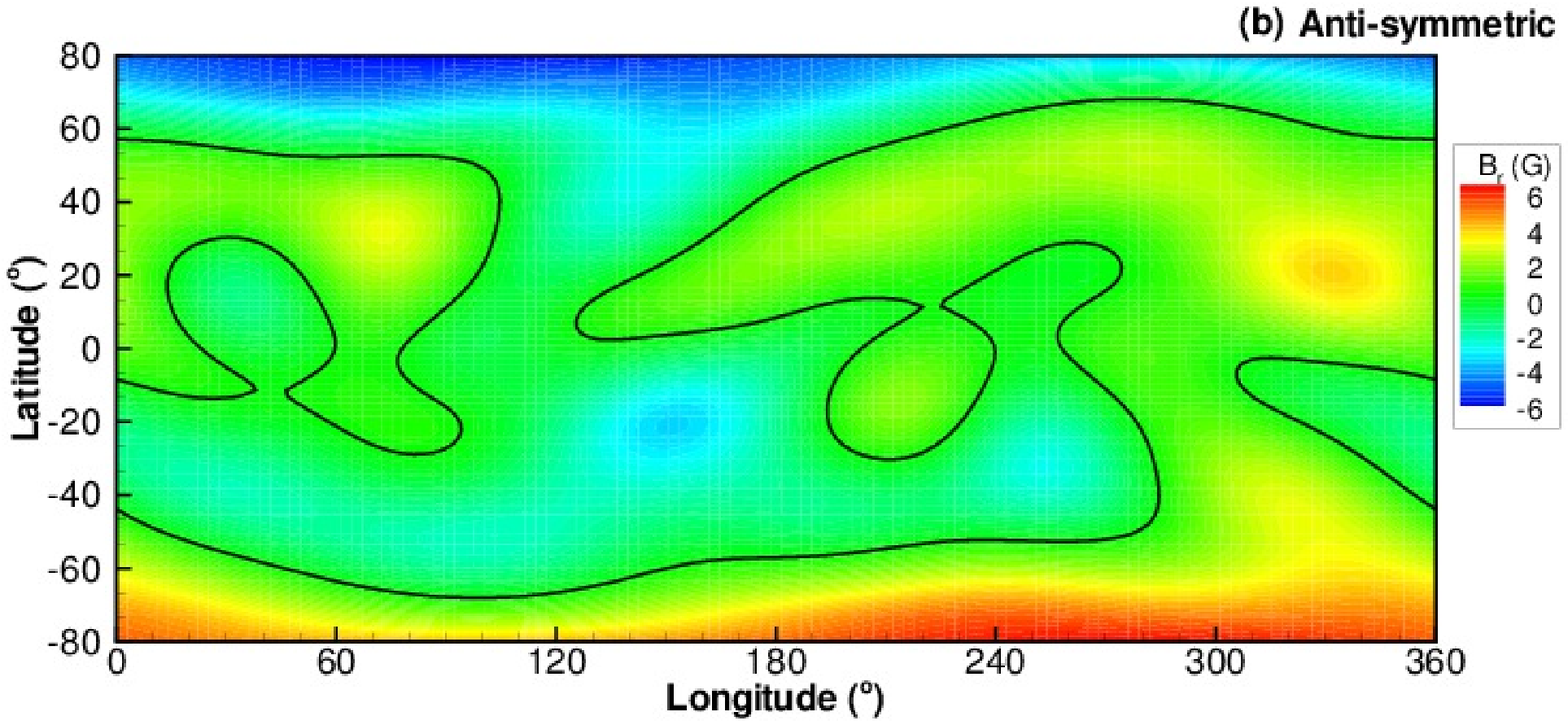}\\
\includegraphics[width=84mm]{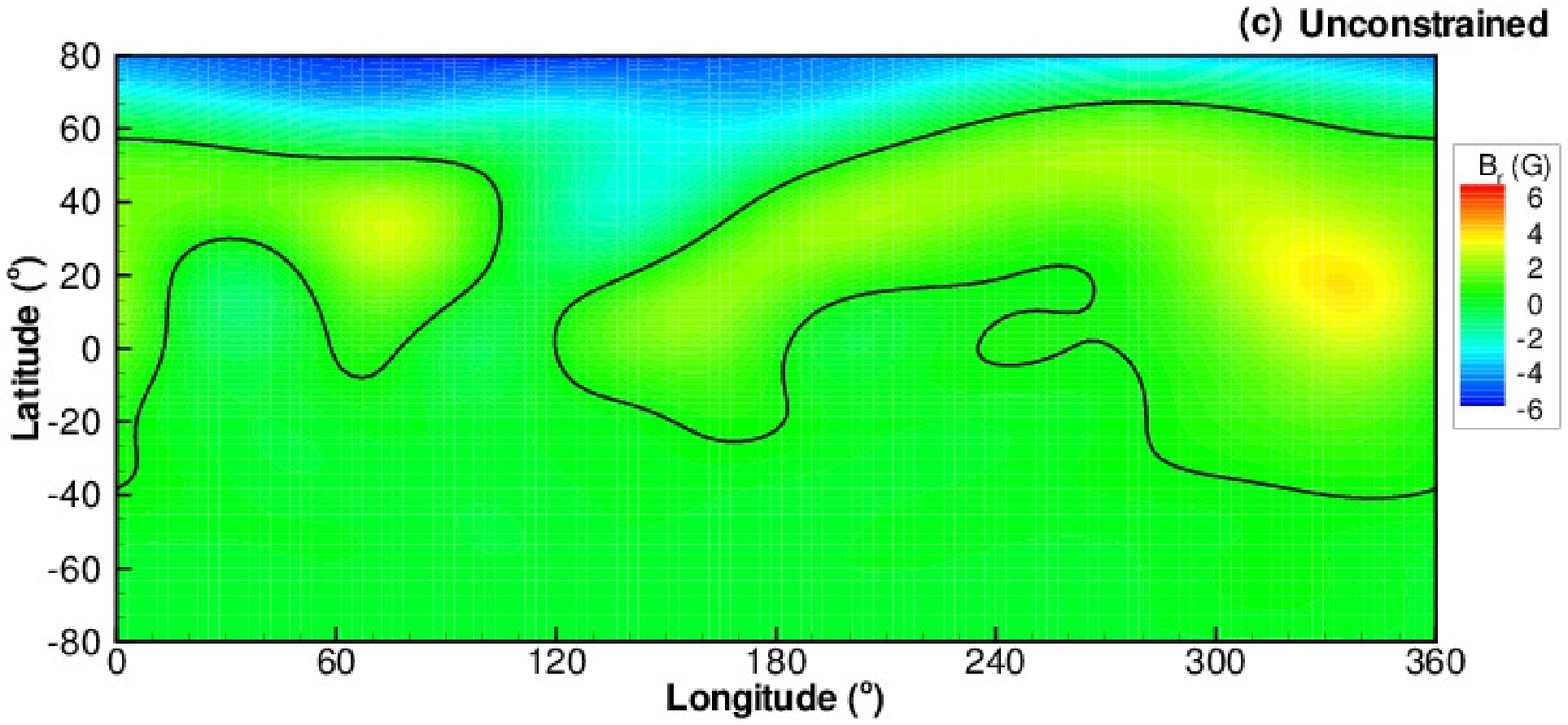}
\caption{Surface distribution of the radial component of the magnetic field reconstructed from observations using ZDI and assuming that (a) the magnetic field is symmetrical with respect to the centre of the star, (b) anti-symmetrical and (c) no constraints are adopted. The investigation shown here is done for the observing epoch of July-2008. The colour-scale of all maps is adjusted to the maximum and minimum values of $B_r$ for the anti-symmetric case.\label{fig.app_maps}}
\end{figure}

The first constraint assumes the magnetic field to be symmetrical with respect to the centre of the star. In that case, the solution obtained in the reconstruction pushes towards even orders of the multipole expansion (e.g., quadrupole). Figure~\ref{fig.app_maps}a shows the surface map that is derived once the symmetrical constraint is adopted. The second constraint adopts an anti-symmetrical magnetic field (Figure~\ref{fig.app_maps}b), such that the solution essentially contains odd orders of the multipole expansion (e.g., dipoles, octopoles).  For comparison, Figure~\ref{fig.app_maps}c shows the reconstructed image without adopting any constrains on the symmetry of the field (the same as shown in Figure~\ref{fig.magnetograms} but with a different colour-scale.). 
As we can see, the reconstructions in the visible hemisphere remain approximately the same, but a different topology arises in the unseen hemisphere.

Using the magnetic maps presented in Figure~\ref{fig.app_maps}, we compute the stellar wind properties in the same way that was described in Section~\ref{sec.results_wind}. Table~\ref{table_constraints} shows the relevant results of our investigation. We note that the unsigned magnetic flux at the surface of the star ($\Pi_0$), a direct output from the observations, is similar for both the symmetrical and anti-symmetrical solutions, which are larger than the unconstrained option by a factor of  $1.7$. Similarly, we find that $\jdot$ and $\Pi_{\rm open}$ obtained are similar for the cases using the symmetrical and anti-symmetrical maps, both of which are comparable to (although larger than) the values obtained using the unconstrained map. 
These similarities suggest that, for the parameters adopted in our model, the choice of the constraints adopted in the reconstruction of the surface magnetic field should not affect the stellar wind results obtained in this paper. 

\begin{table} 
\centering
\caption{Dependence of our results for July-2008 with respect to the assumptions adopted on the symmetrical properties of the field  by the reconstruction method: unconstrained, symmetrical and anti-symmetrical with respect to the centre of the star (column 1). The remaining columns are, respectively: the mass-loss rate ($\mdot$), angular momentum-loss rate ($\jdot$), unsigned surface ($\Pi_0$) and open  ($\Pi_{\rm open}$) magnetic fluxes.
\label{table_constraints}}    
\begin{tabular}{ccccc}  
\hline
Assumption&	$\dot{M}/10^{-12} $	 &	$	\dot{J}	$	& $\Pi_0$ &		$\Pi_{\rm open}$ \\
 &	$	 (\msano)$  &	$ (10^{32} ~\rm{erg})$ &	$(10^{22} ~{\rm Mx})$ &$(10^{22} ~{\rm Mx})$  \\ \hline
Unconst.  & $    2.68   $  & $   1.1   $       &$1.14$    &$    9.0$   \\
Symmetric  &    $    2.67   $    & $   1.4   $        &$1.94$    &$12.2$ \\
Anti-sym.&$    2.69   $      & $   1.4   $        &$1.97$    &$   12.2$ \\  
 \hline
\end{tabular}
\end{table}

\begin{figure}
\includegraphics[width=84mm]{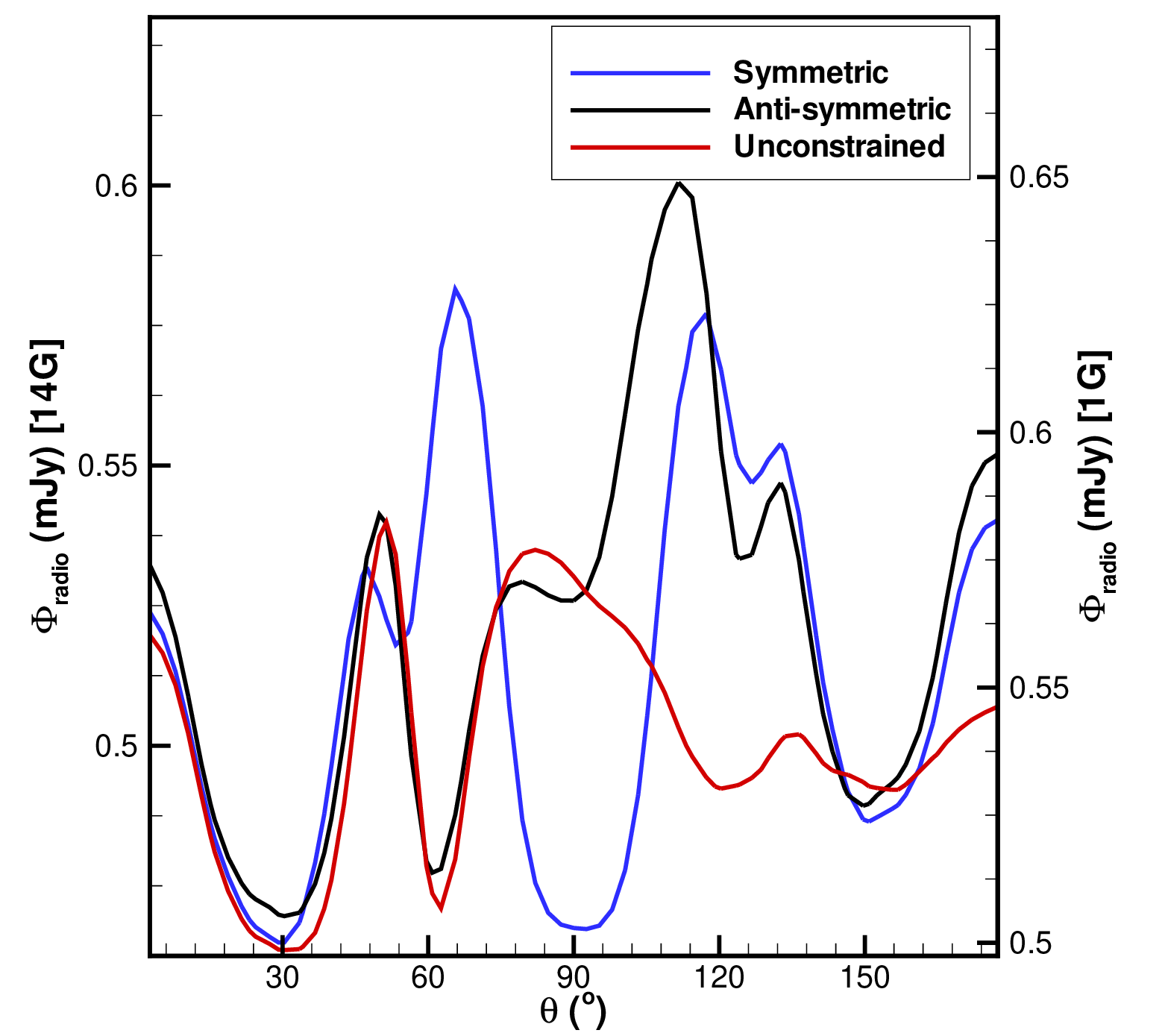}
\caption{Radio flux estimated for $\tau$~Boo b (July-2008) as a function of colatitude $\theta$, which is related to the unknown orbital inclination. The planet is either assumed to have a magnetic field intensity similar to the Earth (right axis) or similar to Jupiter's (left axis). Red solid curve assumes no constraints in the reconstruction of the surface magnetic field (red solid line in Figure~\ref{fig.radio}d), blue solid curve assumes symmetry with respect to the centre of the star, and black solid curve assumes anti-symmetry. \label{fig.app_radio_flux}}
\end{figure}

Using the results of the stellar wind simulations, we proceed to evaluate the exoplanetary radio emission as shown in Section~\ref{sec.radio}. Figure~\ref{fig.app_radio_flux} shows the resultant radio flux emitted by a planet interacting with its host star's wind. The solid lines in Figure~\ref{fig.app_radio_flux} are labeled according to the assumption involved in the reconstruction of the stellar surface magnetic field. Because in the unconstrained case, the reconstructed magnetic field in the unseen hemisphere has smaller intensities and is less structured than that in the visible hemisphere, the stellar wind in the unseen region is less influenced by the latitude-dependent magnetic forces and, therefore, is more spherical. As a consequence, the exoplanetary radio emission are relatively `flat' for a range of colatitudes from $\theta \sim 120^{\rm o}$ onwards  (note that we omitted the region with $\theta \gtrsim 120^o$ in Figures~\ref{fig.theta_shock} and \ref{fig.radio} due to the lack of information there). When the symmetric or anti-symmetric assumptions are imposed, the more complicated topology of the magnetic field is reflected in the radio flux calculated for $120^{\rm o} \lesssim \theta \leq 180^{\rm o}$, as can be seen in Figure~\ref{fig.app_radio_flux} (compare blue and black lines against red one). In spite of that, there is no significant difference between the calculated radio fluxes (symmetric: $0.46-0.58$~mJy, anti-symmetric: $0.47-0.60$~mJy, unconstrained: $0.46-0.59$~mJy). This suggests that the choice of the constraints adopted in the reconstruction of the surface magnetic field should not affect the results obtained in this paper.

\bsp


\label{lastpage}

\end{document}